\documentclass[10pt,aps,prl,twocolumn,superscriptaddress,floatfix,longbibliography]{revtex4-1}

\usepackage{graphicx}
\usepackage{amssymb}
\usepackage{amsmath}
\usepackage{amsfonts}
\usepackage{txfonts}
\usepackage{lipsum}
\usepackage{dcolumn}
\usepackage{xcolor}
\usepackage{bm}
\usepackage{enumitem}
\usepackage{wasysym}
\usepackage{bbold}
\usepackage[colorlinks=true, allcolors=blue]{hyperref}
\usepackage{mathtools}
\usepackage{hhline}
\usepackage{soul}
\usepackage{ulem}

\begin{document}

\title{Local and nonlocal electronic correlations at the metal-insulator transition\\
in the Hubbard model in two dimensions}

\author{Maria Chatzieleftheriou}
\affiliation{CPHT, CNRS, \'Ecole polytechnique, Institut Polytechnique de Paris, 91120 Palaiseau, France}

\author{Silke Biermann}
\affiliation{CPHT, CNRS, \'Ecole polytechnique, Institut Polytechnique de Paris, 91120 Palaiseau, France}
\affiliation{Coll\`ege de France, 11 place Marcelin Berthelot, 75005 Paris, France}
\affiliation{European Theoretical Spectroscopy Facility, 91128 Palaiseau, France}

\author{Evgeny A. Stepanov}
\affiliation{CPHT, CNRS, \'Ecole polytechnique, Institut Polytechnique de Paris, 91120 Palaiseau, France}
\affiliation{Coll\`ege de France, 11 place Marcelin Berthelot, 75005 Paris, France}

\begin{abstract} 
Elucidating the physics of the single-orbital Hubbard model in its
intermediate coupling regime is a key missing ingredient to our understanding of
metal-insulator transitions in real materials. 
Using recent non-perturbative many-body techniques that are able to interpolate
between the spin-fluctuation-dominated Slater regime at weak coupling and the
Mott insulator at strong-coupling, we obtain the momentum-resolved spectral function 
in the intermediate regime and disentangle the effects of antiferromagnetic 
fluctuations and local electronic correlations in the formation of an insulating state. 
This allows us to identify the Slater and Heisenberg regimes in the phase diagram, 
which are separated by a crossover region of competing spatial and local electronic correlations. 
We identify the crossover regime by investigating the behavior of the local magnetic 
moment, shedding light on the formation of the insulating state at intermediate couplings.

\end{abstract}

\maketitle

Ever since the seminal works by Slater \cite{PhysRev.82.538} and
Mott \cite{Mott}, correlation-induced metal-insulator transitions (MIT)
in transition metal compounds such as vanadates, titanates, cuprates or nickelates have intrigued the 
condensed matter and materials science community. 
The Hubbard model, a single-orbital lattice model describing the interplay of kinetic energy and Coulomb 
interactions, is a minimal model for such a MIT and has become the drosophila of the field. 
While in the limiting cases of one and infinite spatial dimensions the exact solutions of the
Hubbard model are known, the nature of the MIT in finite dimensions (and in particular in the important 
case of the model on the $d=2$ square lattice) is still a subject of intense debate. 
In the weak coupling regime, the half-filled Hubbard model with nearest-neighbour hopping amplitudes
undergoes a N\'eel transition to an antiferromagnetic (AFM) insulating state at a critical temperature $T_N$.
In this Slater regime~\cite{PhysRev.82.538, PhysRevB.94.125144}, the formation of the insulating
state is driven by Fermi surface (FS) nesting, enabling fluctuations between FS points that are connected 
by the AFM wave vector. 
In the opposite -- strong-coupling -- limit, strong local Coulomb interaction localizes the electrons, inducing
a Mott insulating state~\cite{Mott, RevModPhys.70.1039} with Heisenberg-type magnetic moments~\cite{PhysRev.115.2, CHAO1977163, Chao_1977, PhysRevB.37.9753, spalek2007tj}.
Upon decreasing temperature, at $T_N$ a Mott insulator undergoes a paramagnetic-to-AFM transition, 
which is no longer driven by the FS nesting, as the FS is absent in the Mott phase.
Close to the paramagnetic Mott transition, the theoretical challenge consists 
in solving, at least approximately, the quantum many-body problem in the intermediate coupling regime
that cannot be reached by perturbative expansions around the weak-coupling (band) or strong-coupling
(localized) limits. Moreover, in the proximity of the phase transition long correlation lengths are 
expected, calling for techniques going beyond the local picture of dynamical mean field theory (DMFT) 
\cite{RevModPhys.68.13} methods.

Recent years have seen the development of a flury of attempts in this direction, including
\textit{cluster} methods, where spatial (rather short-range) correlations are taken into account within 
a cluster of lattice sites~\cite{PhysRevB.58.R7475, PhysRevB.62.R9283, PhysRevLett.87.186401, RevModPhys.77.1027, doi:10.1063/1.2199446, RevModPhys.78.865, PhysRevLett.101.186403, PhysRevB.94.125133, Cluster_Fratino} and \textit{diagrammatic} methods, where nonlocal correlations are treated via a diagrammatic expansion around DMFT~\cite{RevModPhys.90.025003, Lyakhova_review}, such as $GW$+DMFT~\cite{PhysRevLett.90.086402, PhysRevLett.92.196402, PhysRevLett.109.226401, PhysRevB.87.125149, PhysRevB.90.195114, PhysRevB.94.201106, PhysRevB.95.245130}, dual fermions~\cite{PhysRevB.77.033101, PhysRevB.79.045133, PhysRevLett.102.206401, PhysRevB.94.035102, PhysRevB.96.035152, BRENER2020168310}, dual bosons~\cite{Rubtsov20121320, PhysRevB.90.235135, PhysRevB.93.045107, PhysRevB.94.205110, Stepanov18-2, PhysRevB.100.165128, PhysRevB.102.195109}, the dynamical vertex approximation~\cite{PhysRevB.75.045118, PhysRevB.80.075104, PhysRevB.95.115107, doi:10.7566/JPSJ.87.041004, PhysRevB.103.035120}, the triply irreducible local expansion (TRILEX)~\cite{PhysRevB.92.115109, PhysRevB.93.235124, PhysRevB.96.104504, PhysRevLett.119.166401}, or the dual TRILEX (\mbox{D-TRILEX})~\cite{PhysRevB.100.205115, PhysRevB.103.245123, 10.21468/SciPostPhys.13.2.036, PhysRevLett.127.207205, stepanov2021coexisting, PhysRevResearch.5.L022016, PhysRevLett.129.096404, Vandelli} methods.
In the weak coupling regime, diagrammatic Monte Carlo~\cite{PhysRevLett.81.2514, Kozik_2010, PhysRevLett.119.045701, PhysRevB.100.121102, PhysRevLett.124.117602} studies
of the Hubbard model on the 2d square lattice \cite{PhysRevX.11.011058,PhysRevX.5.041041, PhysRevB.91.235114}
provide the numerically exact solution, evidencing characteristic momentum-selective effects stemming from 
AFM fluctuations as expected in the Slater regime.
In the intermediate coupling regime, however, and despite considerable progress in recent years,
there is still no complete picture of the physics of even the single-orbital model at half-filling.

In this Letter, we use the recently developed \mbox{D-TRILEX} method to elucidate the spectral properties
close to and the nature of the Mott transition in the half-filled Hubbard model
on a 2d square lattice.
We analyse corrections to the momentum-resolved spectral functions from nonlocal fluctuations beyond DMFT,
disentangling the contribution of AFM fluctuations and local electronic correlations.
We find the weak- and strong-coupling limits to be separated by a crossover regime that starts when the local magnetic moment (LMM) is formed and ends at the Mott transition.
In this regime, spatial fluctuations are important, but emergent LMM herald strong local electronic correlations in the Mott phase. Increasing $U$ enhances the value of the LMM and concomitantly reduces the momentum-selectivity 
of spectral features on the FS until it eventually disappears upon entering the Mott phase. 
Our work bridges the weak and strong coupling pictures of the MIT
by providing a complete analysis of the interplay of nesting, local moment physics and Mott localisation
in the 2d Hubbard model.

\begin{figure*}[t!]
\includegraphics[width=0.25\linewidth]{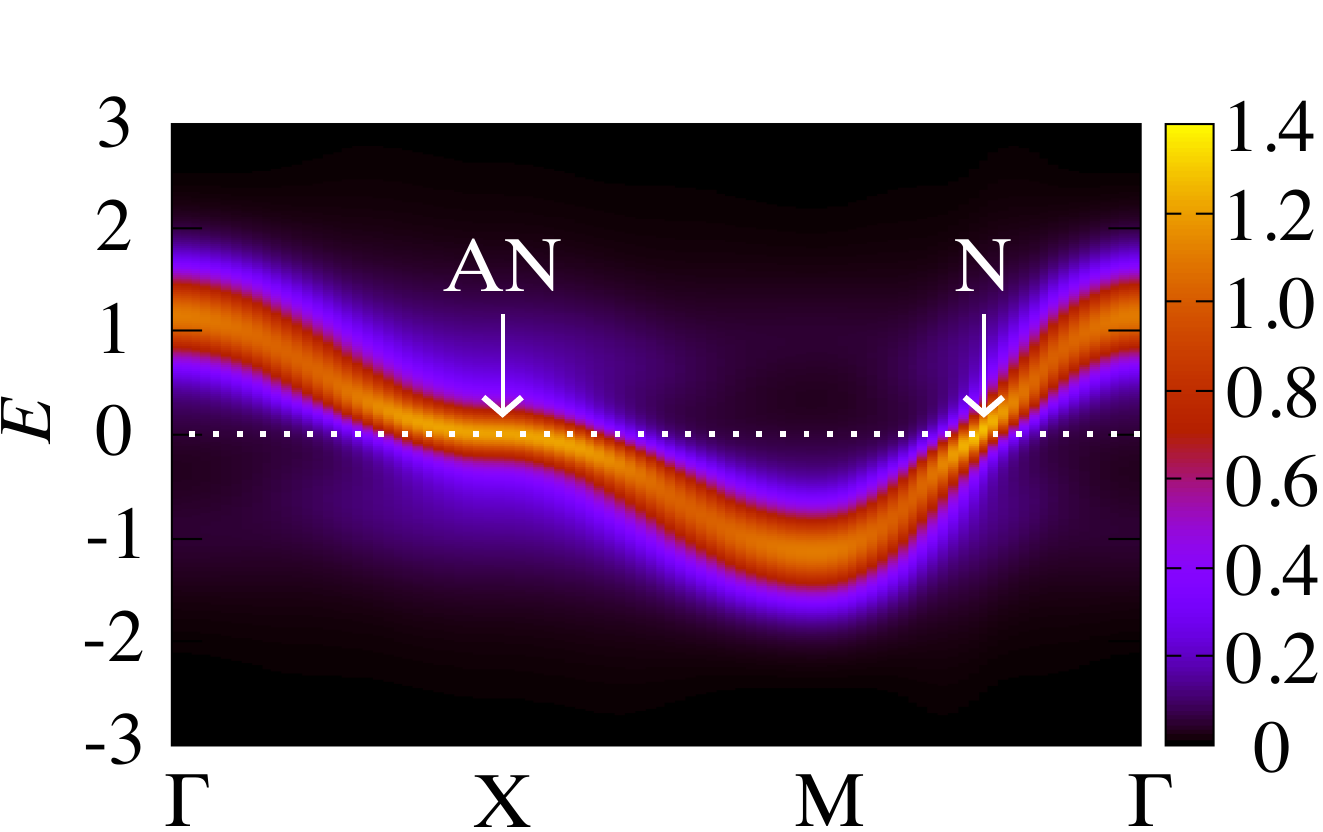}
~~
\includegraphics[width=0.25\linewidth]{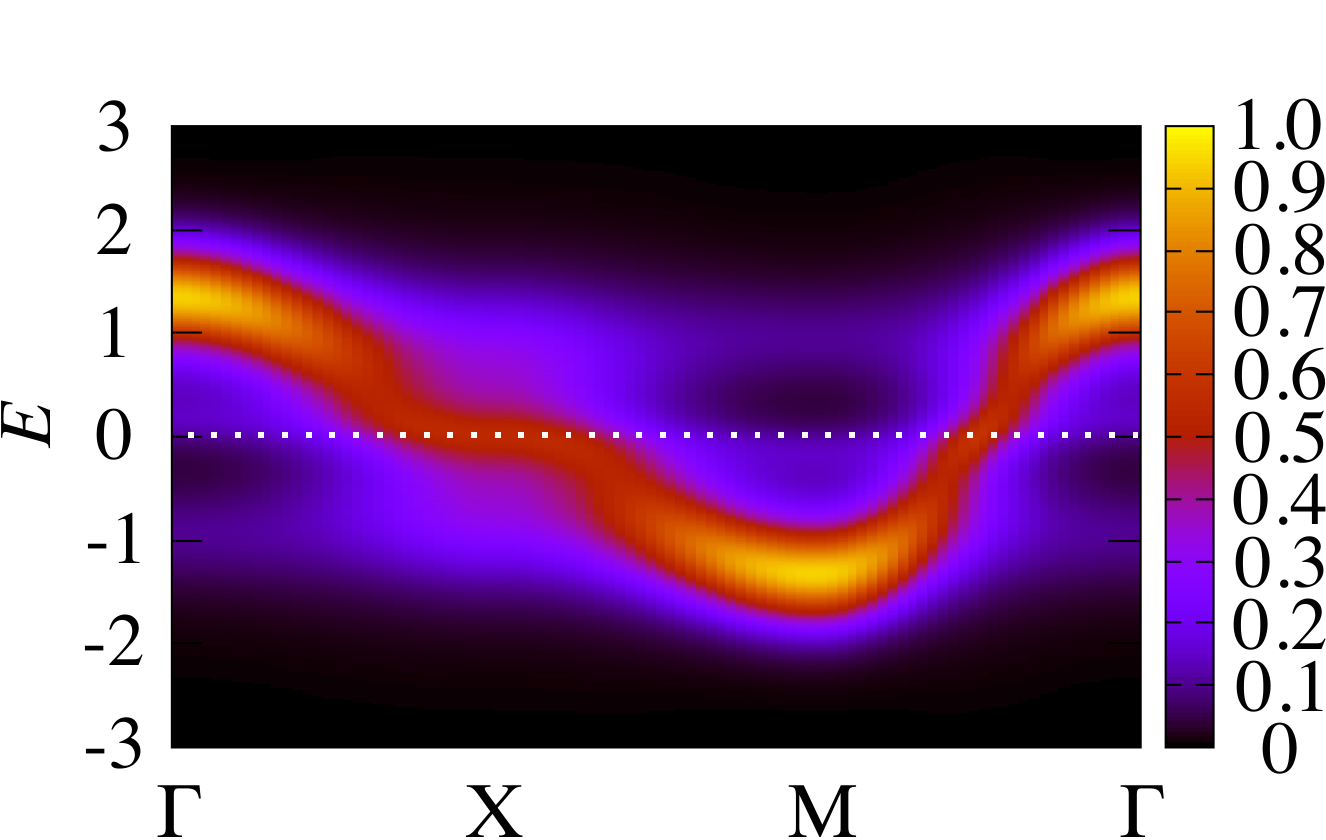}
~~
\includegraphics[width=0.25\linewidth]{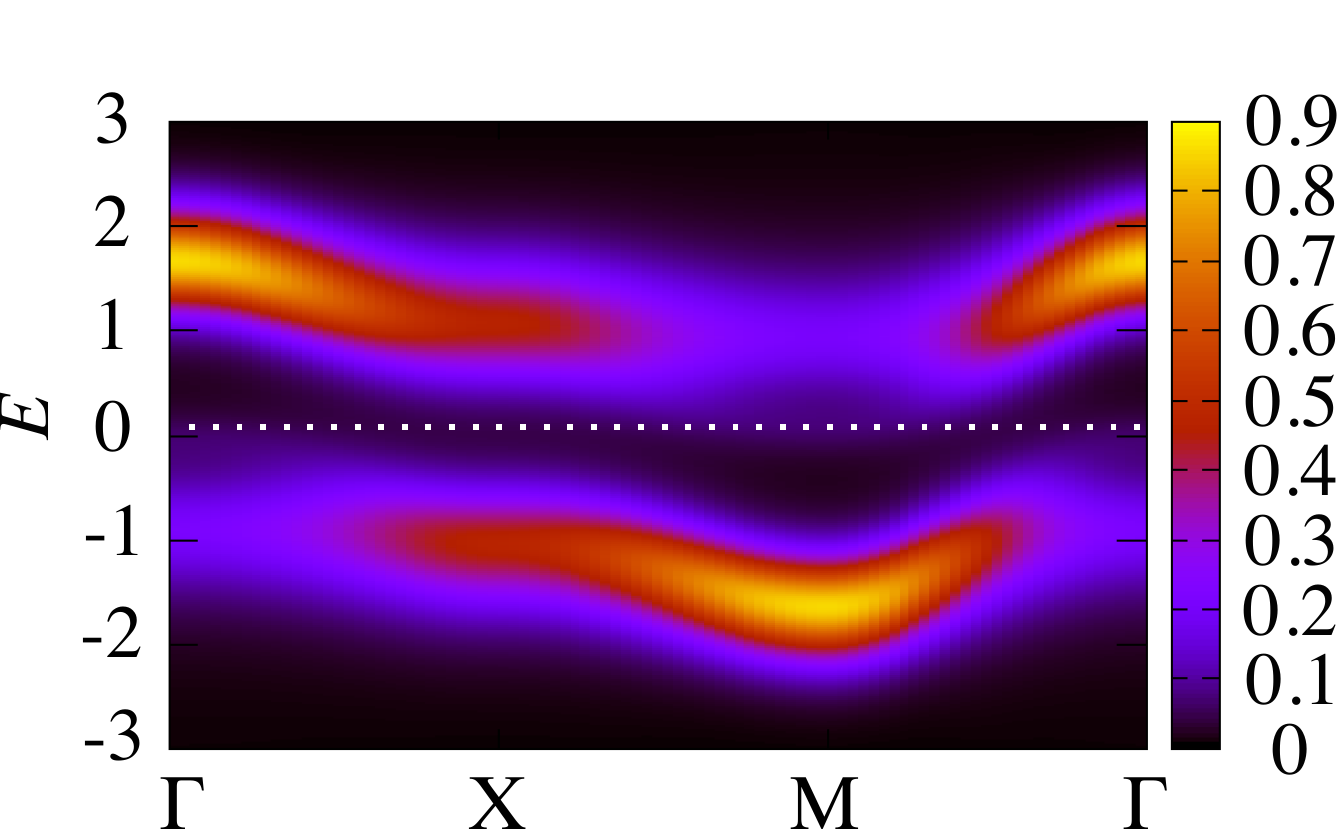}
~~
\includegraphics[width=0.19\linewidth]{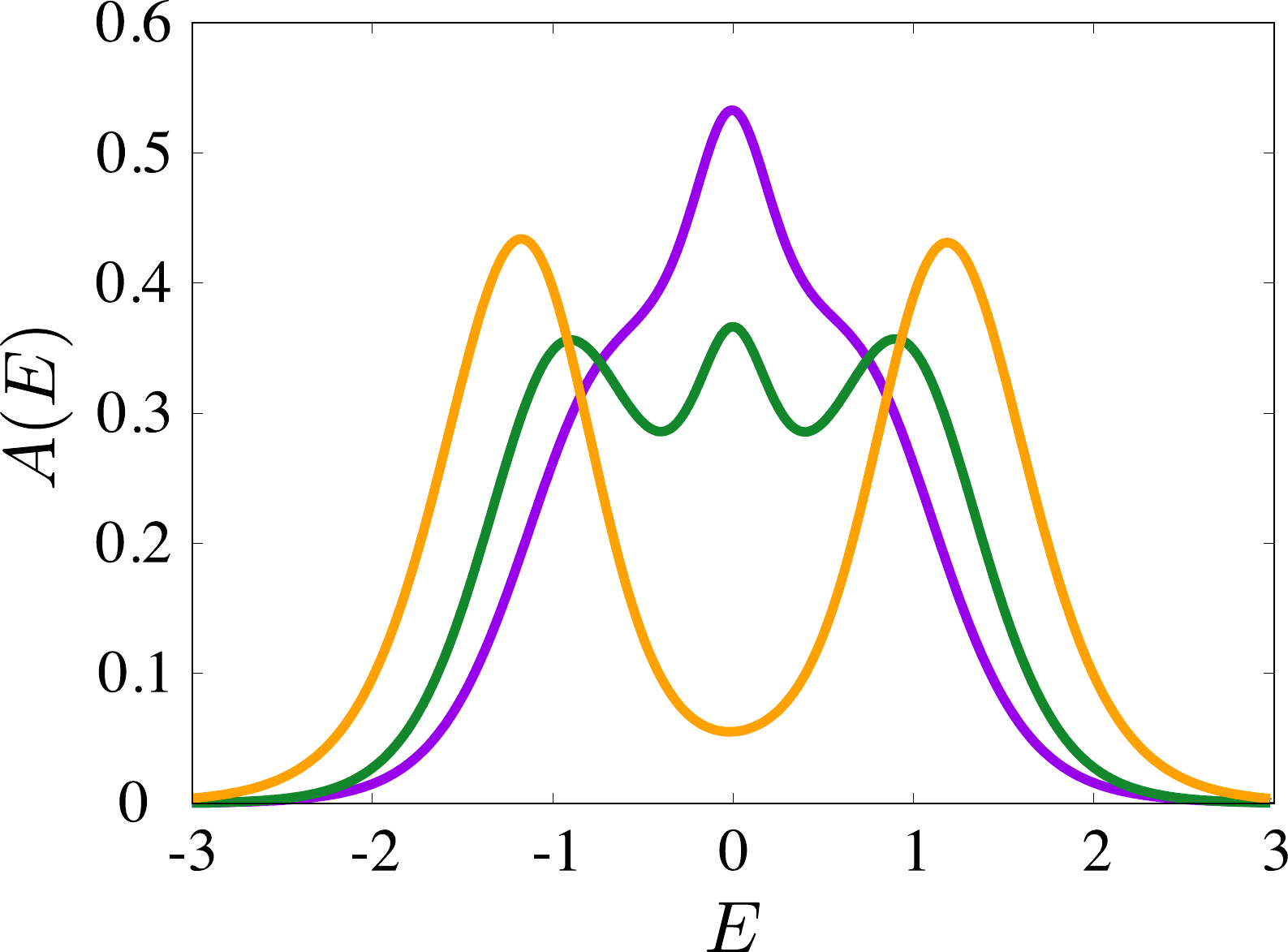}
\\
\includegraphics[width=0.25\linewidth]{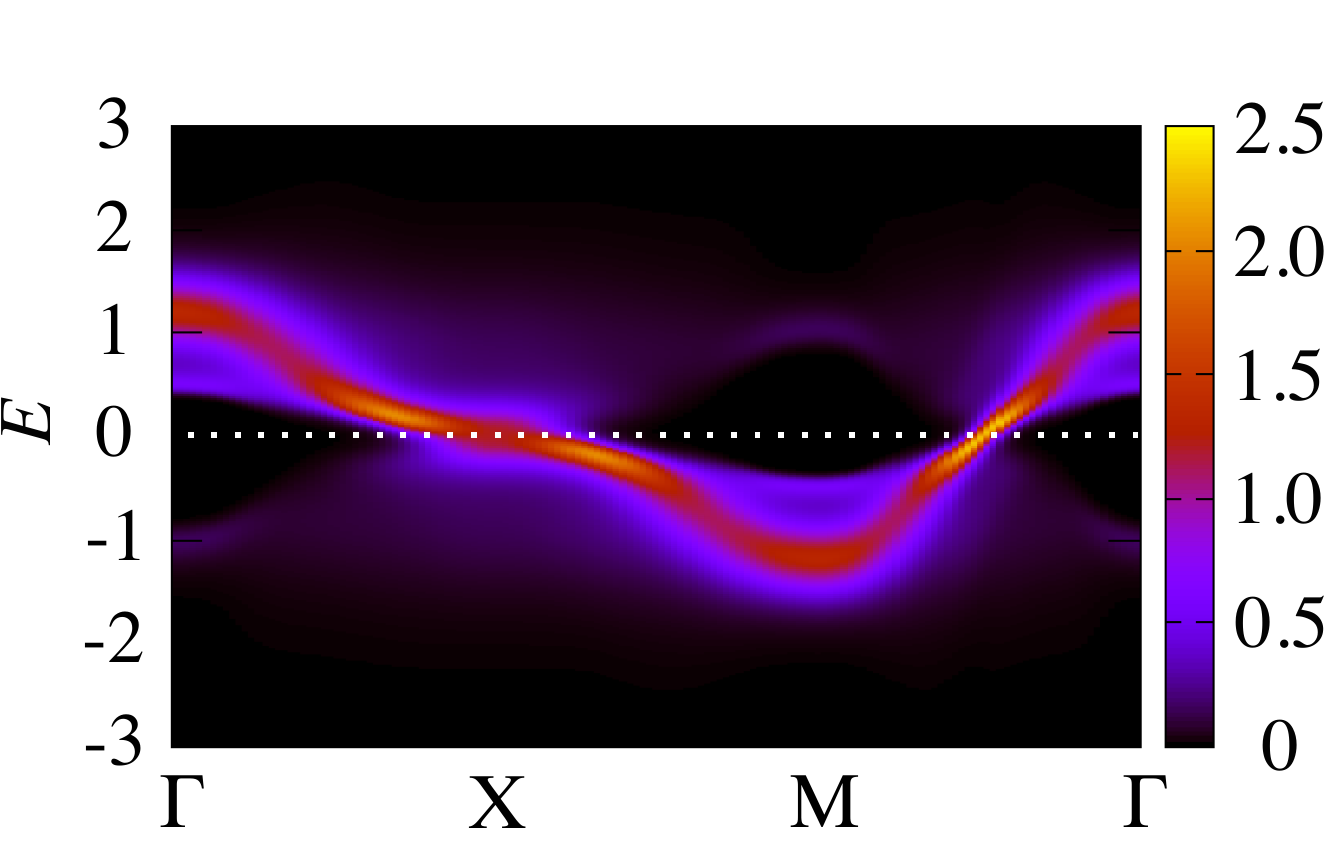}
~~
\includegraphics[width=0.25\linewidth]{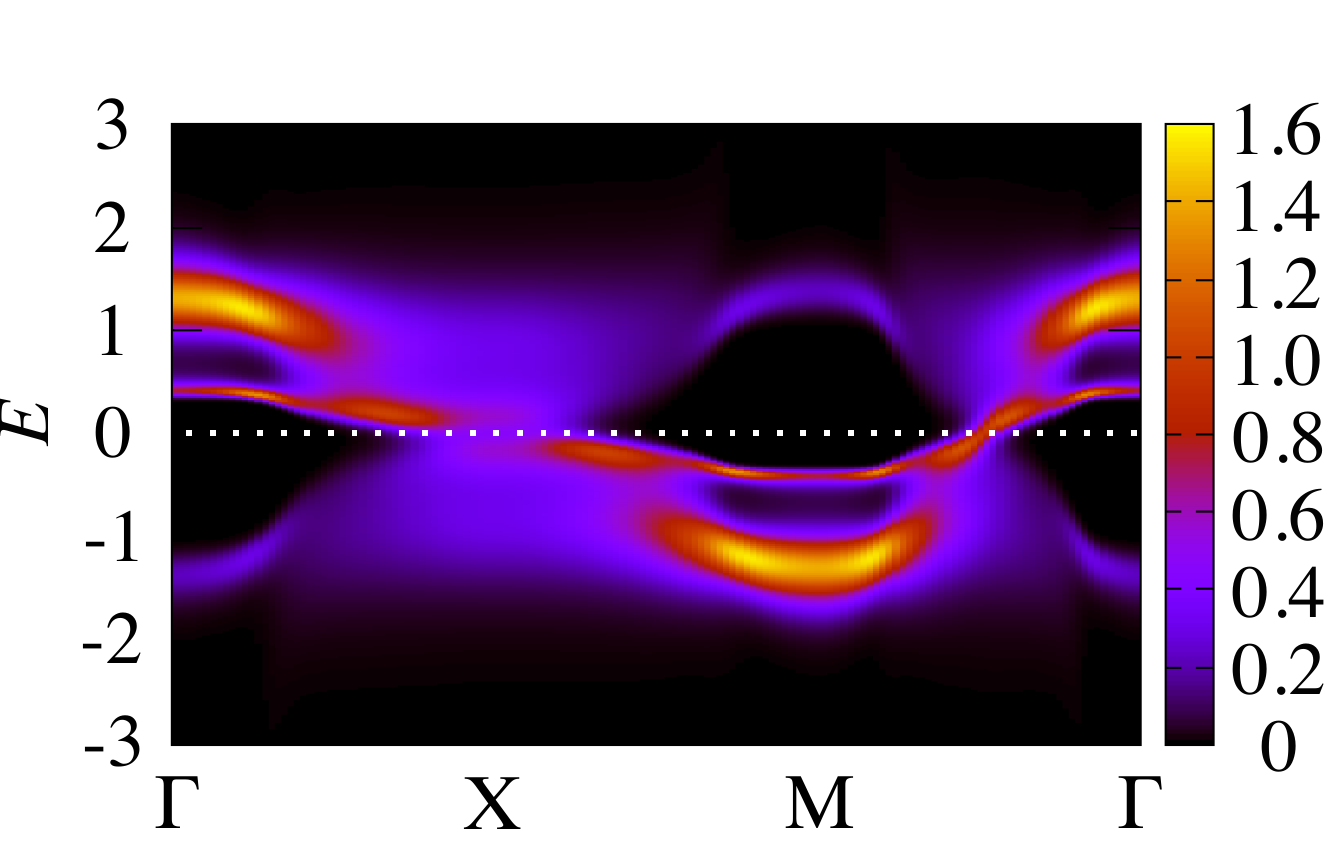}
~~
\includegraphics[width=0.25\linewidth]{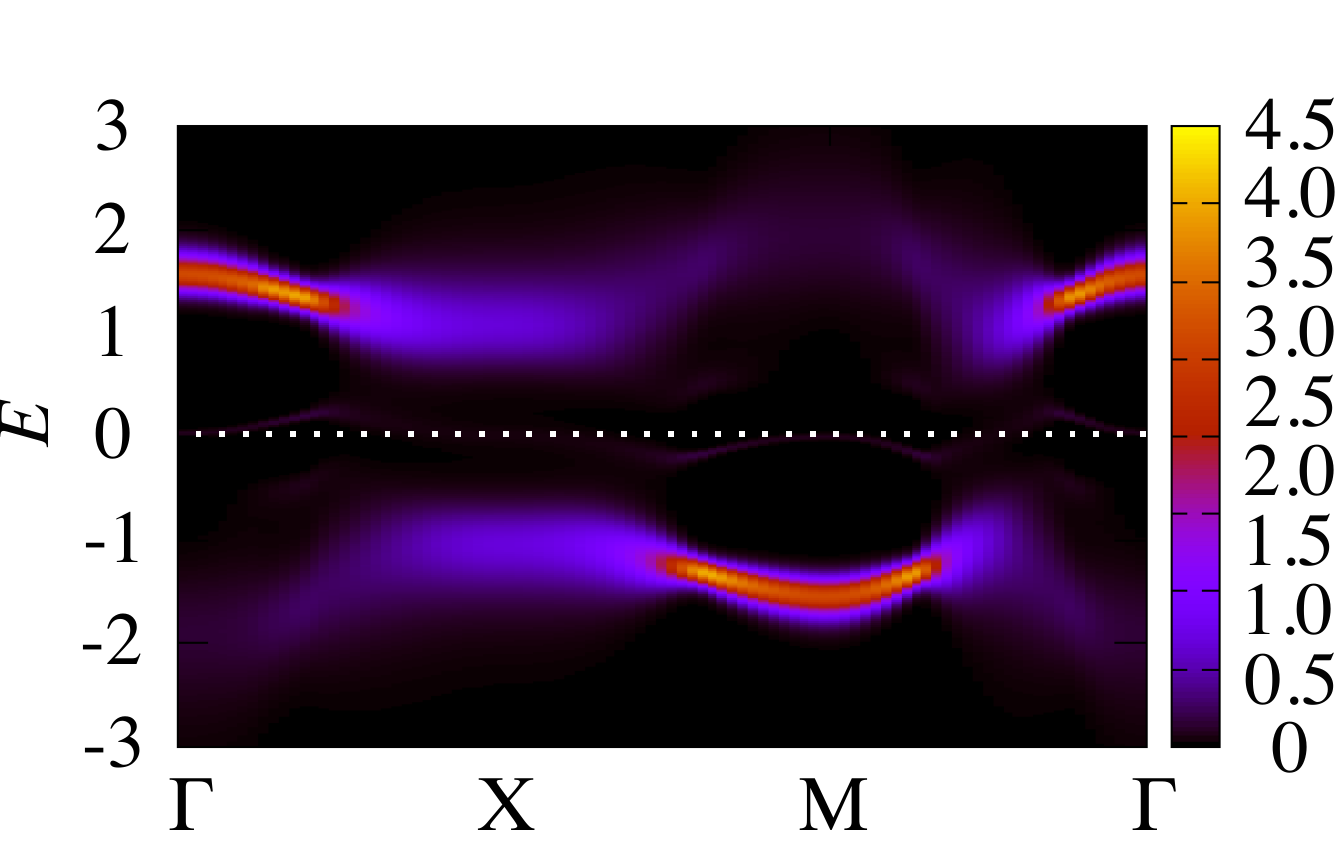}
~~
\includegraphics[width=0.19\linewidth]{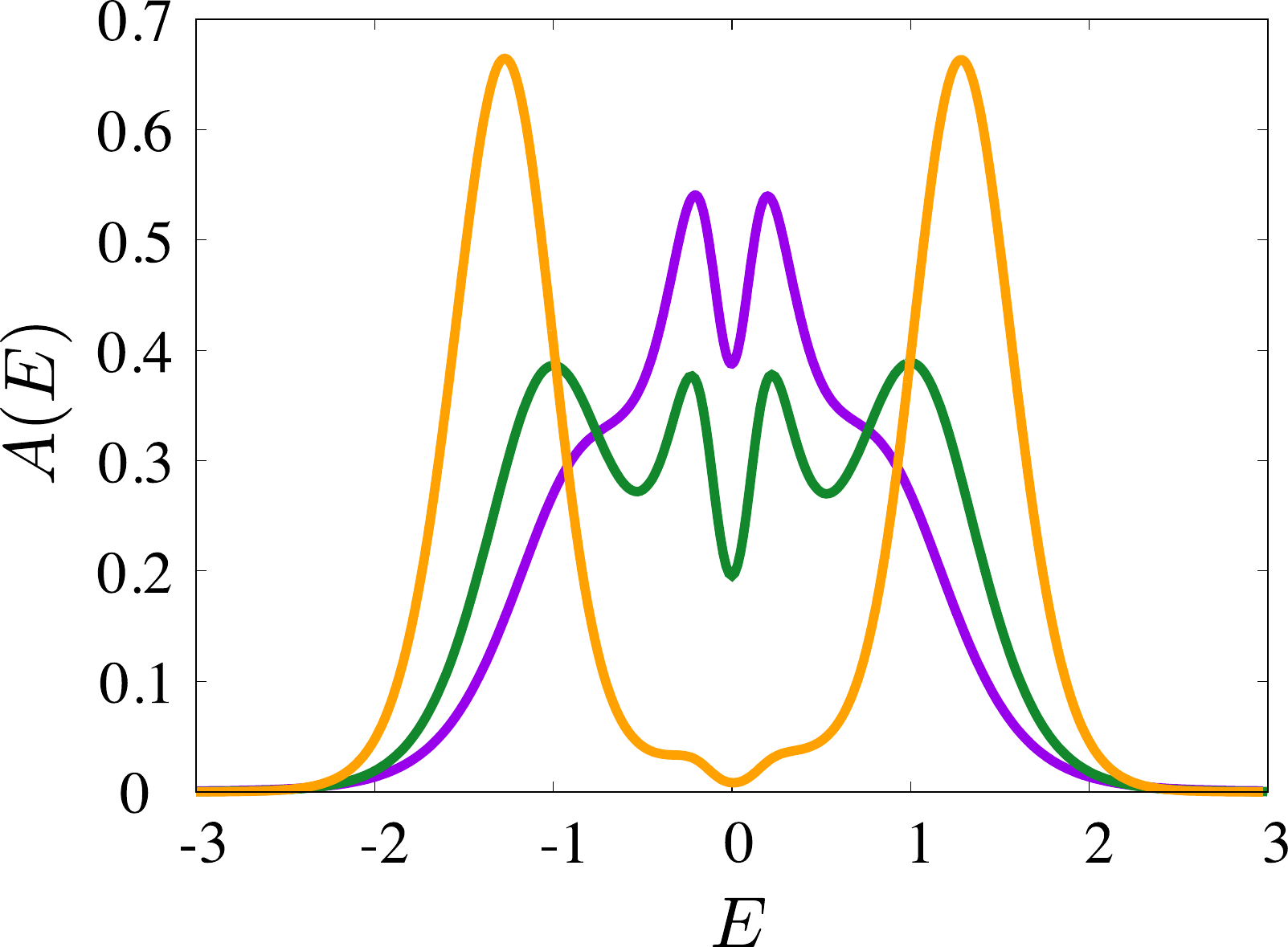}
\caption{Momentum-resolved spectral function of the Hubbard model at half filling,
at $T=0.14$ (upper row) and $T=0.06$ (lower row) within \mbox{D-TRILEX}, at
$U=1.2$ (left), $U=1.6$ (middle) and $U=2.4$ (right),
along the $k$-path $\Gamma-X-M-\Gamma$ with $\Gamma=(0,0)$, $X=(\pi, 0)$, and $M=(\pi, \pi)$.
The far right column displays the corresponding local spectral functions obtained for ${U=1.2}$ (purple), ${U=1.6}$ (green) and ${U=2.4}$ (orange).}
\label{fig:spectr_dtrilex}
\end{figure*}

We study the single-orbital half-filled Hubbard model~\cite{hubbard1963electron}:
${\hat{H} = t\sum_{\langle jj' \rangle, \sigma} c_{j\sigma}^{\dagger} c^{\phantom{\dagger}}_{j'\sigma} + U\sum_{i} n_{j\uparrow} n_{j\downarrow}}$,
where $U$ is the on-site Coulomb interaction. 
We consider a 2d square lattice with dispersion ${\epsilon_{\bf k}=2t(\cos k_x+\cos k_y)}$.
We use the half bandwidth as our unit of energy, i.e. ${t=0.25}$. 
Here, $c^{(\dagger)}_{\sigma}$ is the annihilation (creation) operator for an electron with spin ${\sigma\in\{\uparrow, \downarrow\}}$ and ${n_{j\sigma} = c^{\dagger}_{j\sigma}c^{\phantom{\dagger}}_{j\sigma}}$.

We solve the problem using the \mbox{D-TRILEX} approach~\cite{PhysRevB.100.205115, PhysRevB.103.245123, 10.21468/SciPostPhys.13.2.036}, which allows for a self-consistent treatment of spatial charge and spin fluctuations of arbitrary range~\cite{PhysRevLett.127.207205, stepanov2021coexisting, PhysRevResearch.5.L022016, PhysRevLett.129.096404, Vandelli} that are taken into account diagrammatically beyond DMFT.
The momentum-resolved spectral function ${A(\bf k, \omega) = - \frac{1}{\pi} \rm{Im}\,G(\bf k, E)}$ and its momentum-integrated (local) counterpart ${\rho(\omega) = \sum_{\bf k} A(\bf k, \omega)}$ are obtained by performing analytical continuation for the Matsubara electronic Green's function $G(\bf k, \nu)$ using the maximum entropy method implemented in the ana\_cont package~\cite{kaufmann2021anacont}.
The spin susceptibility in \mbox{D-TRILEX} is given by a Bethe-Salpeter equation (BSE)~\cite{10.21468/SciPostPhys.13.2.036} which, in 
the single-band case, reduces to the Dyson equation. It diverges when the ``leading eigenvalue'' $\lambda$ of BSE
approaches $1$. We will use the value of $\lambda$ as a proxy for antiferromagnetic (AFM) fluctuations. 
To determine the role of nonlocal electronic correlations we compare the \mbox{D-TRILEX} results to the 
paramagnetic DMFT solution of the problem (given in the Supplementary Material (SM)~\cite{SM}).

Fig.~\ref{fig:spectr_dtrilex} displays the {\bf k}-resolved spectral function along a high symmetry path in the Brillouin zone (BZ)
and its local counterpart in the high- (top row) and low-temperature (bottom row) regimes for three values of the interaction.
At the high temperature, with increasing interaction strength -- from left to right -- the system evolves 
from a weakly correlated metal, with a spectral function resulting from the non-interacting
one by mere temperature broadening, to a Mott insulator, where the Mott gap results from a splitting
of the spectral features into upper and lower Hubbard bands, separated by the Hubbard interaction ${\sim{}U}$, as expected.
In spite of the gap being well visible in the momentum-resolved spectral function, in the local
spectral function, the strong thermal fluctuations at this temperature lead merely to a minimum
of the spectral weight at the Fermi energy ($E_{F}$), rather than vanishing spectral weight.
Results for these same parameters within DMFT are plotted in the SM~\cite{SM}.
The comparison of the \mbox{D-TRILEX} and DMFT results 
allows for a direct assessment of the role of nonlocal electronic correlations.
As expected, at high temperatures, differences between \mbox{D-TRILEX} and DMFT spectra are minor. 
Nonlocal magnetic fluctuations are
negligible in this regime, as also confirmed by the spin leading eigenvalue being 
significantly smaller than ${\lambda=1}$ (typically here ${\lambda<0.5}$)).
In this regime, the MIT is thus driven by purely local Mott physics and DMFT is sufficient for addressing 
the problem. 

The picture changes radically, when cooling down. While within DMFT temperature-induced
changes of the spectral function are minor and amount to a mere reduction of the width of the different
spectral features, \mbox{D-TRILEX} unravels strongly temperature-dependent changes
in the spectra.
At the largest interaction studied, vanishing spectral weight at the Fermi level confirms the opening of
the Mott gap, both in DMFT and \mbox{D-TRILEX}. Nevertheless, even at this relatively strong interactions,
qualitative differences are observed between the purely local picture of DMFT and the \mbox{D-TRILEX} result.
The differences in the local spectral functions can be traced
back to the differences in the k-resolved ones (see bottom row in Fig.~\ref{fig:spectr_dtrilex}).
We observe, in particular, an overestimation of the widths of the Hubbard bands within DMFT, while
\mbox{D-TRILEX} displays sharper excitations. 
Furthermore, within \mbox{D-TRILEX}, a ``mirroring'' phenomenon of the electronic dispersion with 
respect to zero energy sets in. 
Such a mirroring results from
the proximity of the system to an ordered AFM state. Indeed, band backfoldings are a trivial effect
occuring when the unit cell is doubled by some kind of ordering, e.g. of AFM nature.
Here, they appear in a non-trivial way, from AFM fluctuations in the paramagnetic phase. 
Remarkably, within \mbox{D-TRILEX} the mirroring of the bands is obtained purely diagrammatically, as 
calculations are performed using a single-site reference problem without any doubling of
the unit cell. The strong AFM fluctuations are
encoded in the nonlocal parts of the self-energy. 
We rationalize this behaviour in the SM~\cite{SM}. 
The mirroring of the bands is noticeable already at ${U=1.2}$ (bottom left panel in 
Fig.~\ref{fig:spectr_dtrilex}) when the fluctuations are already relatively strong 
(${\lambda=0.89}$) and is well pronounced at ${U=1.6}$ (middle panel), 
when the leading eigenvalue of magnetic fluctuations approaches unity (${\lambda=0.94}$).
Further increase of the interaction to ${U=2.4}$ results in the Mott transition, and the mirrored 
dispersion transforms into the two Hubbard bands (right panel).

\begin{figure}[t!]
\includegraphics[width=\linewidth]{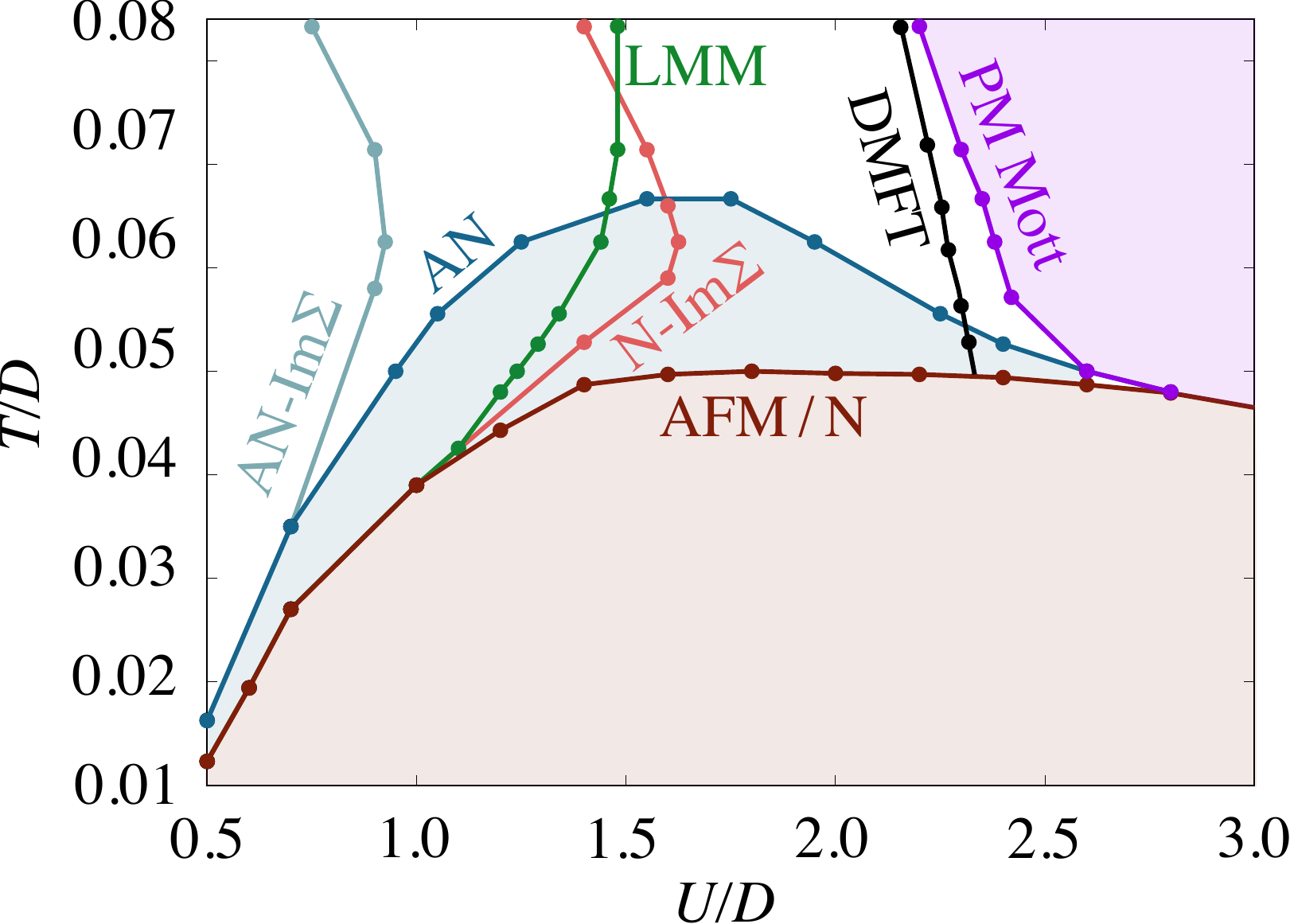}
\caption{Low temperature phase diagram of the 2d Hubbard model within \mbox{D-TRILEX}. 
The different lines have been determined as follows:
Divergence of the spin susceptibility at ${{\bf q}=(\pi,\pi)}$ (dark red line). Sign change of the slope of the self-energy at the AN point (light blue curve). Gap opening in the spectral function at the AN point (dark blue curve).
Sign change of the slope of the self-energy at the N point (light red curve). Formation of non-zero local
magnetic moment (green curve).
The purple line corresponds to the simultaneous disappearance of the quasi-particle peak at $E_{F}$ for both N and AN points, identified with the paramagnetic Mott transition. The black curve is the equivalent result from single-site DMFT calculations. 
\label{fig:phase_diag_weakcoupling}}
\end{figure}

The spectral functions close to $E_F$ reveal characteristic {\bf k}-dependent features. 
Already at ${U=1.2}$ (bottom left panel) we observe a depletion of the spectral weight around the anti-nodal (${{\rm AN}={\rm X}=(\pi,0)}$) point, while the nodal (N) (${{\rm N}=(\pi/2,\pi/2)}$) point
remains unaffected. 
In the local spectral function, the reduction of the spectral weight at the AN point results in a minimum at the Fermi level. 
When increasing the interaction strength to ${U=1.6}$ the quasi-particle (QP) peak at $E_{F}$ completely 
disappears at the AN point, but yet remains at the N point.
This is a well-known effect of AFM fluctuations in the formation of an insulating state, which implies 
that the gap in the spectral function first opens at the AN point (middle panel at the bottom row), 
then propagates along the FS, and finally opens at the N point.
Comparison of the momentum-resolved spectral functions from \mbox{D-TRILEX}, displayed
in the bottom row of Fig.~\ref{fig:spectr_dtrilex}, is suggestive of an interpretation of the
spectral function in the intermediate coupling regime as a superposition of (mirrored) high-energy
Hubbard bands and a low-energy QP band split by AFM fluctuations. 

Inspection of the local spectral functions for the two lowest interaction values in the light
of the k-resolved spectra, calls for an important caveat: obviously, from the depletion of the
local spectral weight at the Fermi level one cannot establish insulating behavior of the system.
Indeed, in the presence of nonlocal correlations, the local spectral function is no longer
the appropriate quantity to look at, since a metallic regime with a momentum-selectively gapped
Fermi surface may not be distinguished from a thermally broadened insulator.

\begin{figure}[t!]
\includegraphics[width=0.48\linewidth]{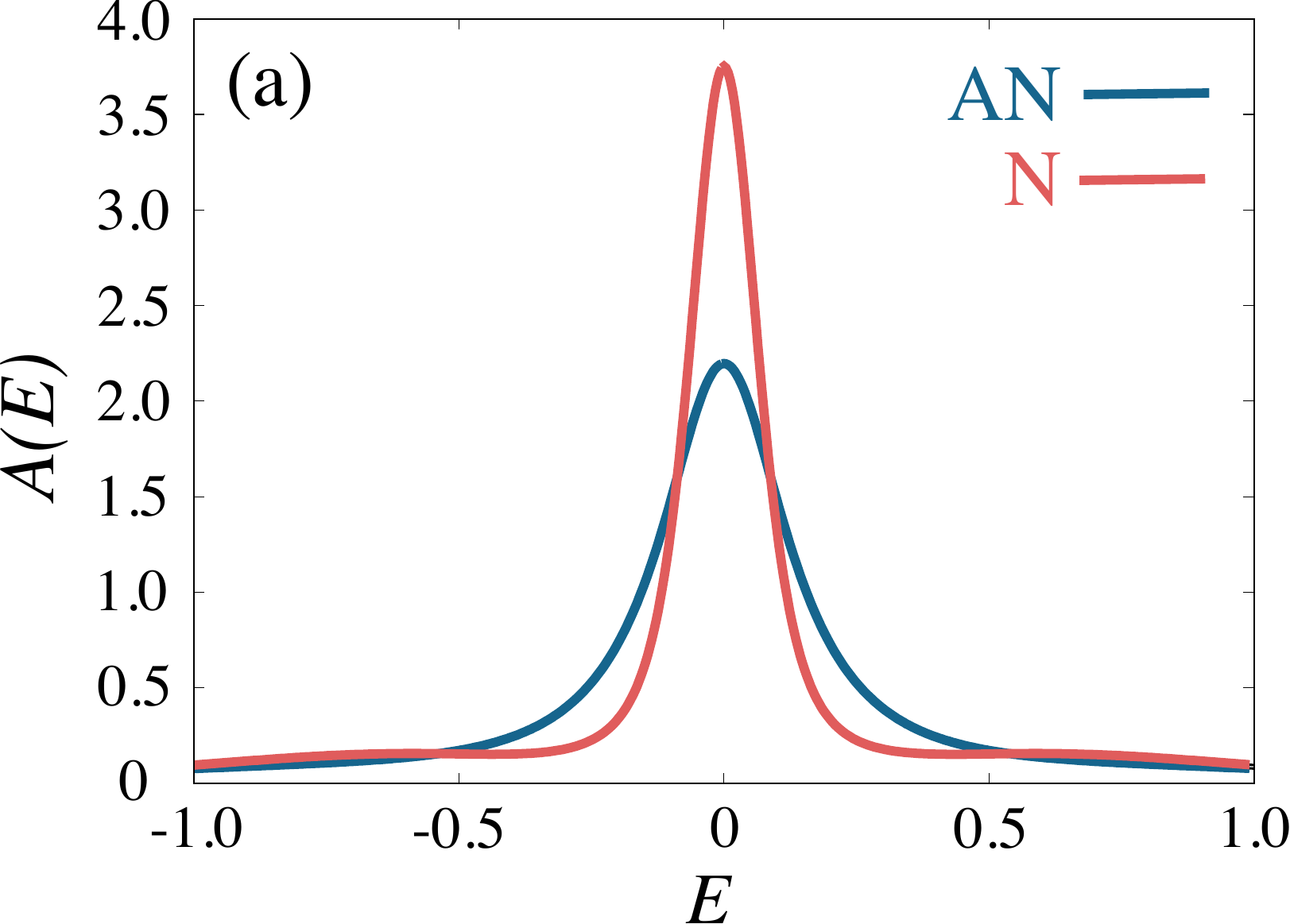}
~~
\includegraphics[width=0.48\linewidth]{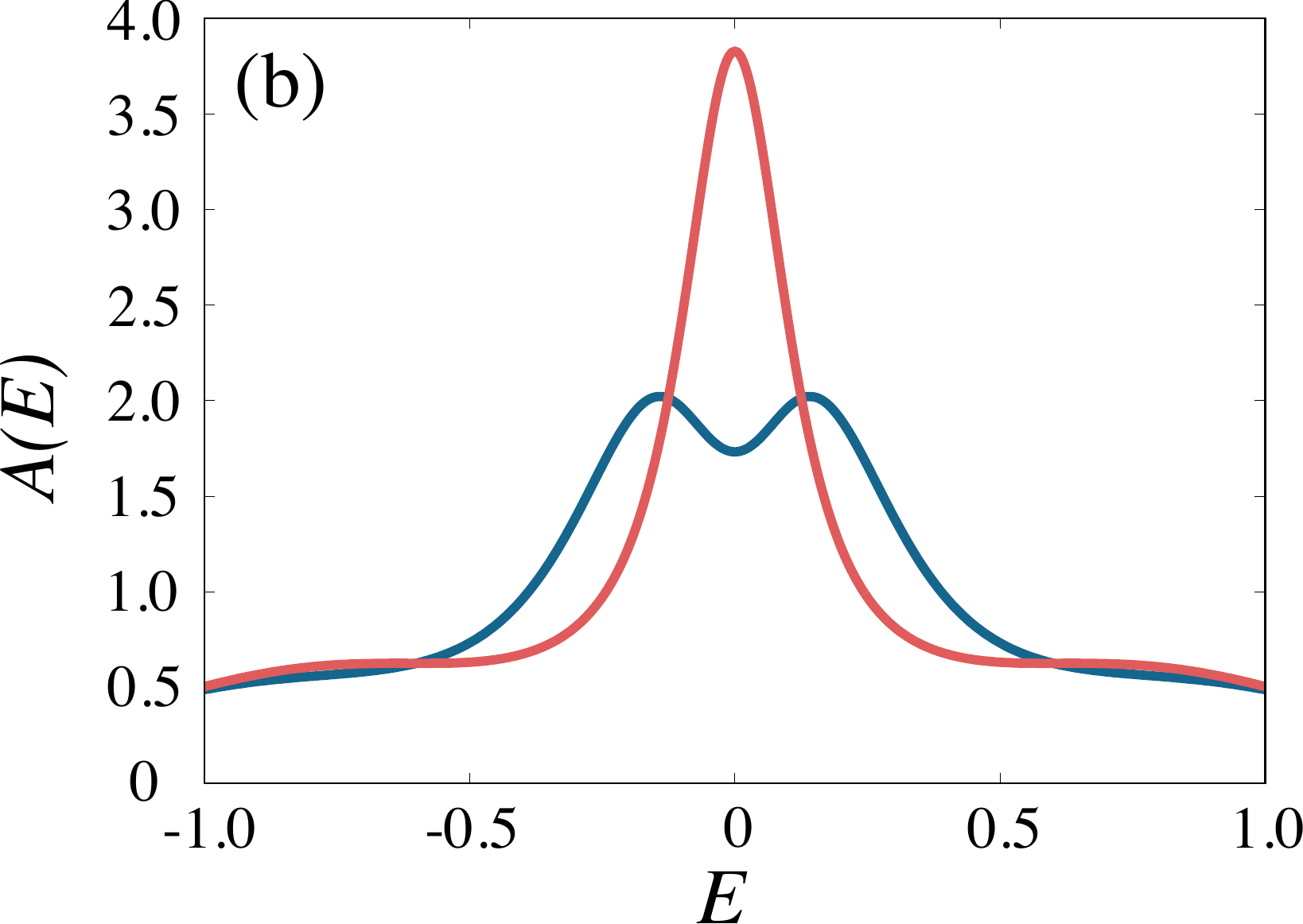}
\\[7pt]
\includegraphics[width=0.48\linewidth]{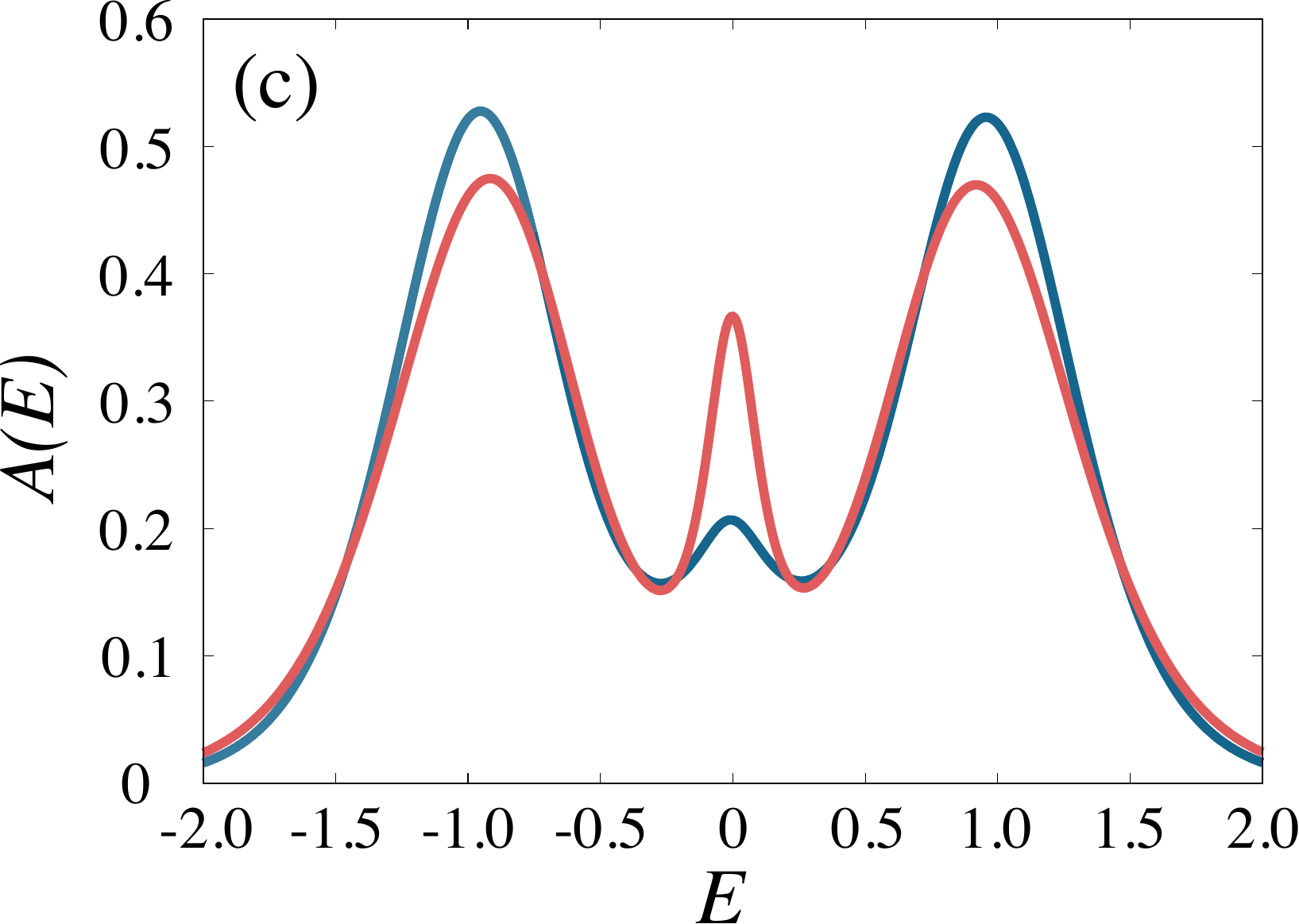}
~~
\includegraphics[width=0.48\linewidth]{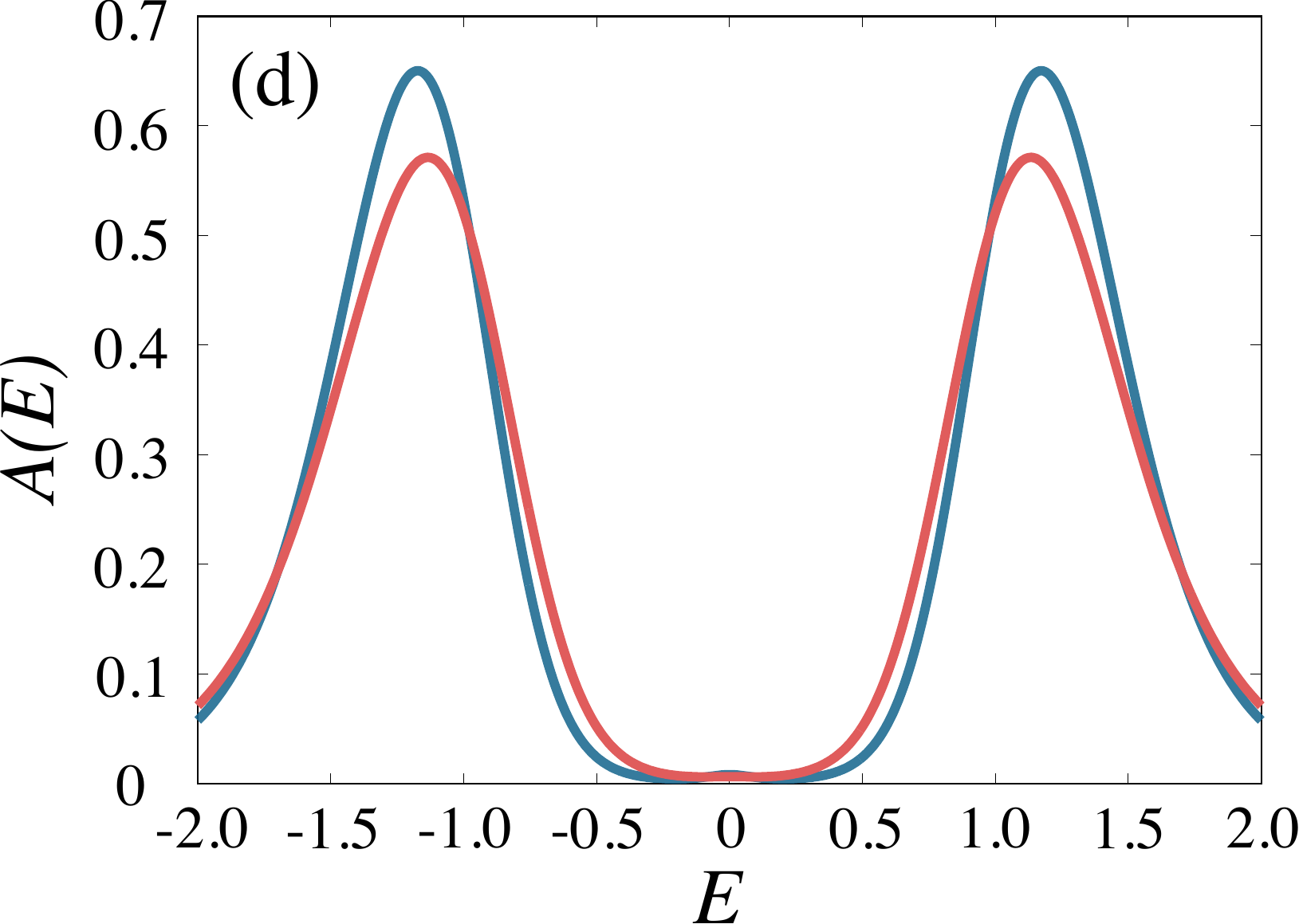}
\caption{Spectral function at the AN (red) and N (blue) points at ${T=0.062}$ for 
${U=1.0}$, ${U=1.4}$, ${U=2.0}$, and ${U=2.6}$. 
For increasing $U$, the QP peak at $E_{F}$ disappears first at the AN point, but is restored at larger values of $U$.  
Finally, the spectra are gapped at both points. 
\label{fig:spectr_N_AN_beta16}}
\end{figure}

The observed momentum-selective disappearance of the quasiparticle peak at $E_{F}$ along the FS 
suggests to revisit the phase diagram and study the spectral function at the AN and N points separately.
In Fig.~\ref{fig:phase_diag_weakcoupling}, we regroup different pieces of information:
The dark red line corresponds to the $(T,U)$-values for which the spin susceptibility at wave vector 
${{\bf q}=(\pi,\pi)}$ diverges, indicating the N\'eel transition to the AFM (quasi-)ordered state at 
lower $T$. 
Concomitantly, in the spectral function, at the N point, the QP peak at $E_F$ is lost,
and a gap opens.
The dark blue line depicts the formation of a minimum at $E_{F}$ in the spectral function at the AN point, which -- as
seen above -- occurs at larger temperatures than at the N point. 
Interestingly, the AN curve (dark blue line) has a dome-like shape mimicking the form of the N\'eel phase boundary.
At ${U\simeq2.8}$ the AN and N curves coincide, and the disappearance of the quasi-particle peak at $E_{F}$ is not momentum-selective any more in the regime of large interactions ${U\gtrsim2.8}$.
Remarkably, the form of the dark blue curve suggests that at certain values of $T$ by {\it increasing} 
the interaction strength $U$ one can move the system from the regime where part of the FS is gapped 
(blue area in Fig.~\ref{fig:phase_diag_weakcoupling}) to a metallic regime (white area between the dark 
blue and purple curves), where QP are restored along the entire FS.
An example is given in Fig.~\ref{fig:spectr_N_AN_beta16}.

Increasing the interaction even more results in a simultaneous disappearance of the QP peak at $E_{F}$ and in the formation of the gap at both AN and N points (Fig.~\ref{fig:spectr_N_AN_beta16}\,d), which is depicted in Fig.~\ref{fig:phase_diag_weakcoupling} by the purple line.
In the low temperature regime the purple curve goes toward larger values of $U$ upon decreasing temperature, until it eventually merges with the AN (dark blue) and N (dark red) lines.
Remarkably, this simultaneous opening of the gap
along the FS is non-momentum-selective and is thus governed by local electronic correlations.
For this reason, we identify the purple curve with the Mott transition. 
Thus, in the regime of weak-to-intermediate interactions AFM fluctuations highly affect the behavior of the system, introducing momentum-selectivity between the N and AN points, which results in the momentum-selective disappearance of the quasi-particle peak along the FS.
This behavior is consistent with a Slater-like scenario for the MIT.
On the contrary, at strong interactions local electronic correlations dominate.

Interestingly, when taking into account nonlocal correlations, 
the paramagnetic Mott transition is actually realized at slightly larger values of $U$ than in single-site DMFT
(see black and purple curves in Fig.~\ref{fig:phase_diag_weakcoupling}).
This is in contrast to results obtained within cluster DMFT methods~\cite{PhysRevLett.101.186403, Cluster_Fratino}, which find a reduction of $U_c$ for the Mott transition compared to single-site DMFT.
This is a combined effect of local and spatial correlations that cannot be easily separated. 
Here we disentangle these two mechanisms and associate the Mott transition to an effect driven by the local correlations. 
An increase of $U_c$ was also discussed using TRILEX~\cite{PhysRevB.93.235124}, based however on an analysis of the
local spectral function at the Fermi energy, which -- for the reasons explained above -- is not a good quantity for determining the MIT.

How does the system interpolate between the two regimes of dominant spatial and local electronic correlations?
The first signature of the crossover between the two regimes is the decrease of the N-AN dichotomy at ${U\gtrsim1.5}$, which manifests itself in a decrease of the difference between the critical temperatures for the disappearance of the quasi-particle peak at $E_{F}$ at the AN and N points (dark blue and red lines in Fig.~\ref{fig:phase_diag_weakcoupling}).
This means that the local electronic correlations start to take over the spatial ones well before the system undergoes the Mott transition.

Also plotted in 
Fig.~\ref{fig:phase_diag_weakcoupling} are the critical values (T,U) 
at which the behavior of the self-energy at the AN (light blue) and N (light red line) points changes from metallic Fermi-liquid (FL)-like to 
incoherent and eventually gapped behavior (dubbed NFL). The self-energies themselves are plotted in the SM~\cite{SM}.
In the weak-coupling regime the dark and light red curves coincide with the disappearance of the QP peak at the N point.
However, at ${U>1}$ and ${T\simeq0.38}$ the light red curve exhibits a turn toward a smaller value of interactions once the temperature increases.
A similar behavior is observed for the light blue curve which starts deviating from the dark blue curve at a similar temperature ${T\simeq0.38}$.
Such a trend has also been detected in the diagrammatic Monte-Carlo study~\cite{PhysRevLett.124.017003, PhysRevLett.124.117602} in the same regime of weak-to-intermediate correlations.

For this regime of increasing interactions, the low frequency behavior of the self-energy is obviously no longer a good 
proxy for the behavior of the system, where local correlations take over. 
In Fig.~\ref{fig:phase_diag_weakcoupling} the formation of the local magnetic moment (LMM),
calculated following Ref.~\onlinecite{PhysRevB.105.155151}, is depicted by the green line. 
In the weak-coupling regime it lies on top of the N\'eel phase boundary (dark red curve), which is consistent
with a Slater mechanism of the N\'eel transition in this regime.
Remarkably, the LMM curve starts deviating from the N\'eel phase boundary at ${U=1.0}$ exactly when the light red line departs from the dark red line. 
This means that at ${U>1.0}$ the MIT transition is no longer determined solely by magnetic fluctuations of itinerant electrons accounted for by the self-energy, as the LMM also starts contributing to the formation of the insulating state.
The formation of the LMM also reduces the momentum-selectivity in the disappearance of the quasiparticle peak at the $E_{F}$. 
Indeed, we find that the LMM curve crosses the AN line at ${U\simeq1.4}$ right before the N-AN dichotomy starts being suppressed. 
This effect can be explained by the fact that the LMM and spatial collective electronic fluctuations are formed by the same electrons. Upon increasing the interaction more electrons are involved in the formation of the LMM and thus less electrons remain for the fluctuations. Eventually, when reaching the critical value, the LMM completely screens the fluctuations and the system enters the Mott insulator regime. 
The formation of the LMM therefore indicates the beginning of the crossover regime separating the weak-coupling Slater part of the phase diagram from the strong-coupling Heisenberg one.
The values of the LMM, given in the SM~\cite{SM}, suggest
that the end of the crossover regime occurs upon reaching the critical value of the LMM effectively when the system undergoes the Mott transition.

In conclusion, we have studied the $T$-$U$ phase diagram of the half-filled single-orbital Hubbard model on a square lattice.
Analyzing the behavior of the momentum-resolved spectral function, we disentangle the contribution of AFM fluctuations and local electronic correlations to the formation of a depletion of spectral weight at the Fermi energy, connecting the weak- and strong-coupling limits.
These two limits are separated by a crossover regime that starts when the local magnetic moment is formed and ends at the Mott transition.
In this regime, spatial fluctuations are still notable, but the presence of a LMM indicates the emergence of strong local electronic correlations in the system.
We identify the weak-coupling region that precedes the crossover with a Slater regime.
There, the MIT is solely governed by AFM fluctuations, which results in a momentum-selective formation of the $E_{F}$ minimum with a pronounced N-AN dichotomy increasing upon increasing interaction. 
In the intermediate coupling (crossover) regime, increasing $U$ enhances the value of the LMM and thus reduces the N-AN dichotomy that eventually disappears upon entering the Mott phase.  
Finally, the Mott phase can be associated with the Heisenberg regime of local magnetic moments.
Indeed, we have found that the Mott transition occurs when the LMM reaches a critical value.
The simultaneous disappearance of the quasi-particle peak along the Fermi surface at the Mott transition  
additionally illustrates the governing role of local electronic correlations for the physics of the system.
As expected, in the Heisenberg regime the N\'eel transition is driven by a decrease in temperature, which separates the Mott phase into paramagnetic and AFM states. 
This work bridges between the weak and strong coupling pictures of the 2d Hubbard model on the square lattice at half filling,
by disentangling the interplay among the different underlying mechanisms.

\begin{acknowledgments}
We thank A.~I.~Lichtenstein and A.~Georges for useful comments and discussions. We acknowledge support from IDRIS/GENCI Orsay under project number A0130901393 and the help of the CPHT computer support team.
\end{acknowledgments}

\bibliography{sample}

\end{document}


\title{Supplemental Material\\[0.5cm]
Local and nonlocal electronic correlations at the metal-insulator transition\\
in the Hubbard model in two dimensions}

\author{Maria Chatzieleftheriou}
\affiliation{CPHT, CNRS, \'Ecole polytechnique, Institut Polytechnique de Paris, 91120 Palaiseau, France}

\author{Silke Biermann}
\affiliation{CPHT, CNRS, \'Ecole polytechnique, Institut Polytechnique de Paris, 91120 Palaiseau, France}
\affiliation{Coll\`ege de France, 11 place Marcelin Berthelot, 75005 Paris, France}
\affiliation{European Theoretical Spectroscopy Facility, 91128 Palaiseau, France}

\author{Evgeny A. Stepanov}
\affiliation{CPHT, CNRS, \'Ecole polytechnique, Institut Polytechnique de Paris, 91120 Palaiseau, France}
\affiliation{Coll\`ege de France, 11 place Marcelin Berthelot, 75005 Paris, France}

\maketitle

\onecolumngrid

\begin{figure}[h!]
\includegraphics[width=0.25\linewidth]{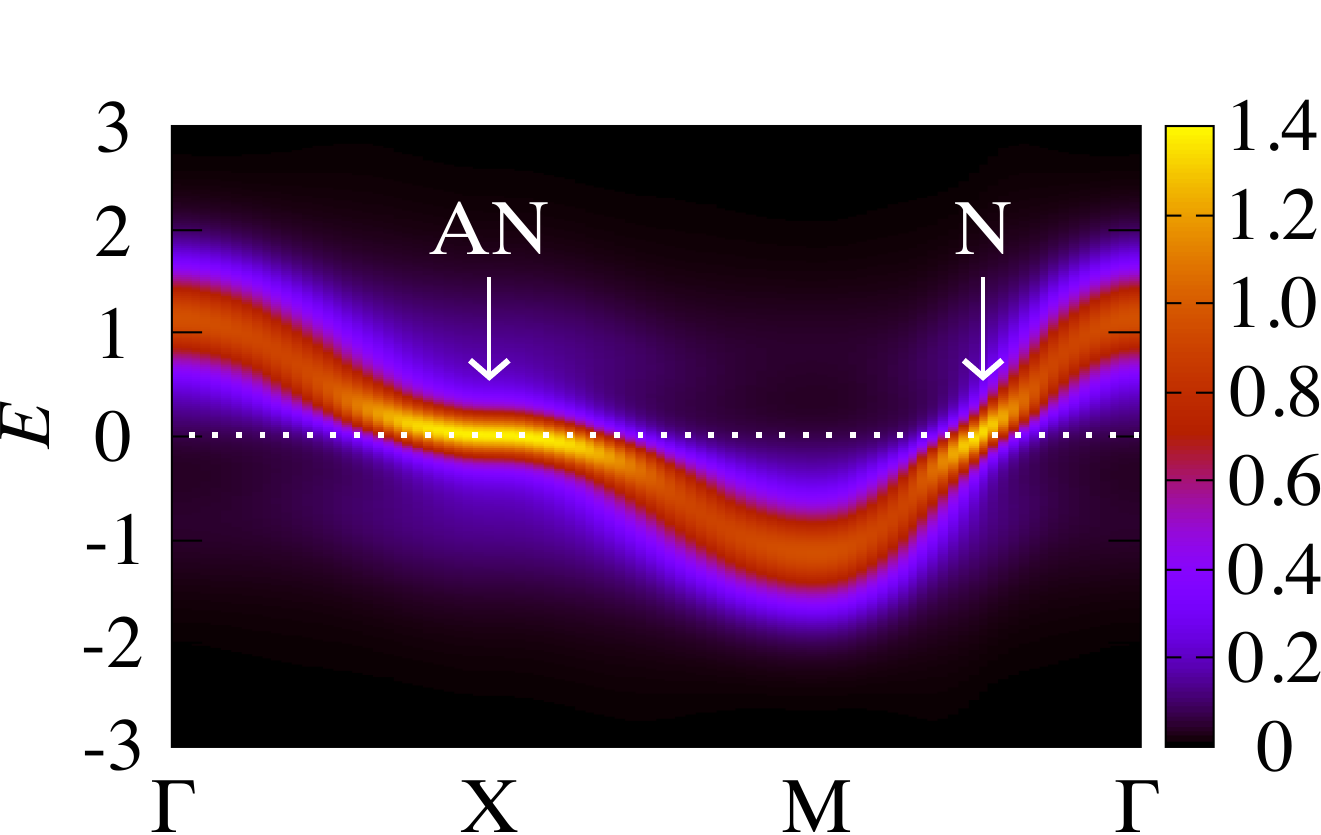}
~~
\includegraphics[width=0.25\linewidth]{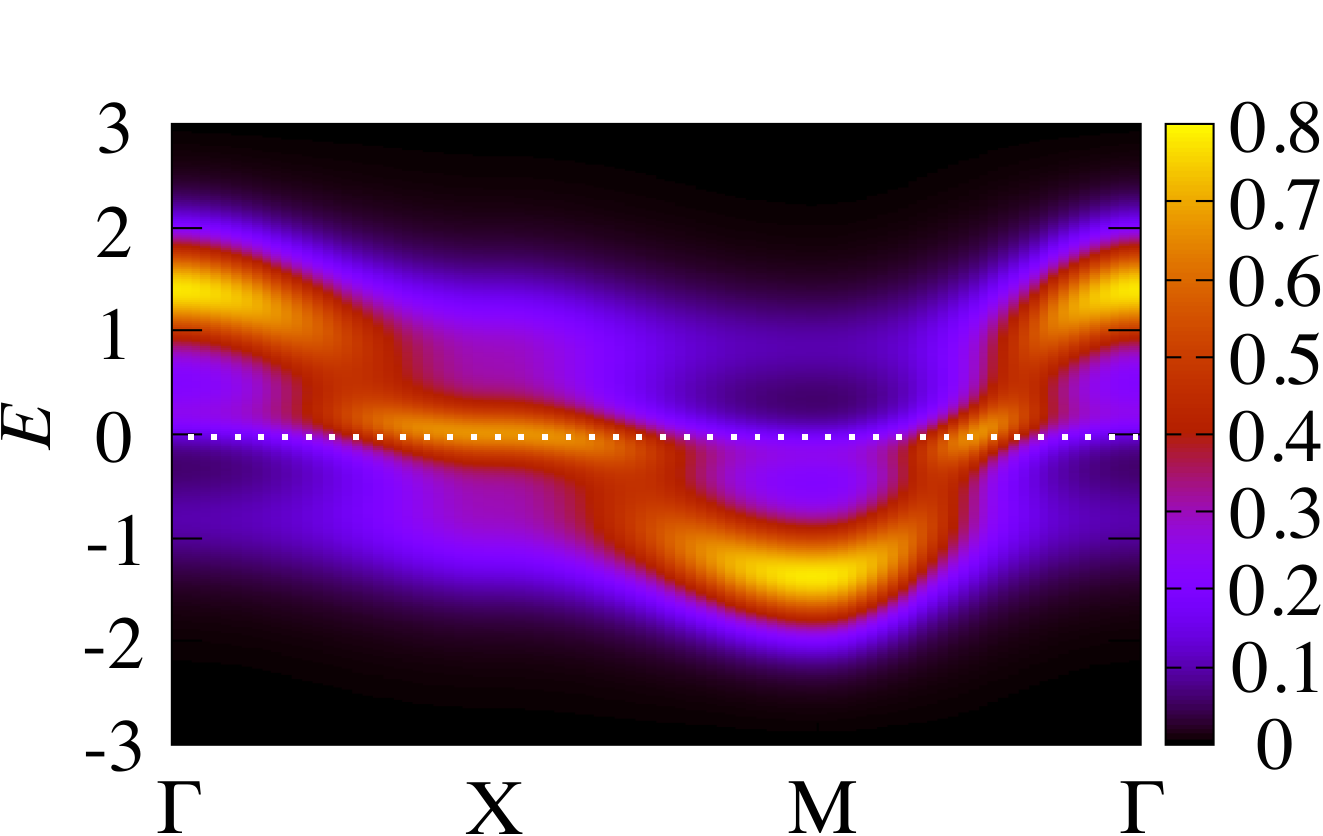}
~~
\includegraphics[width=0.25\linewidth]{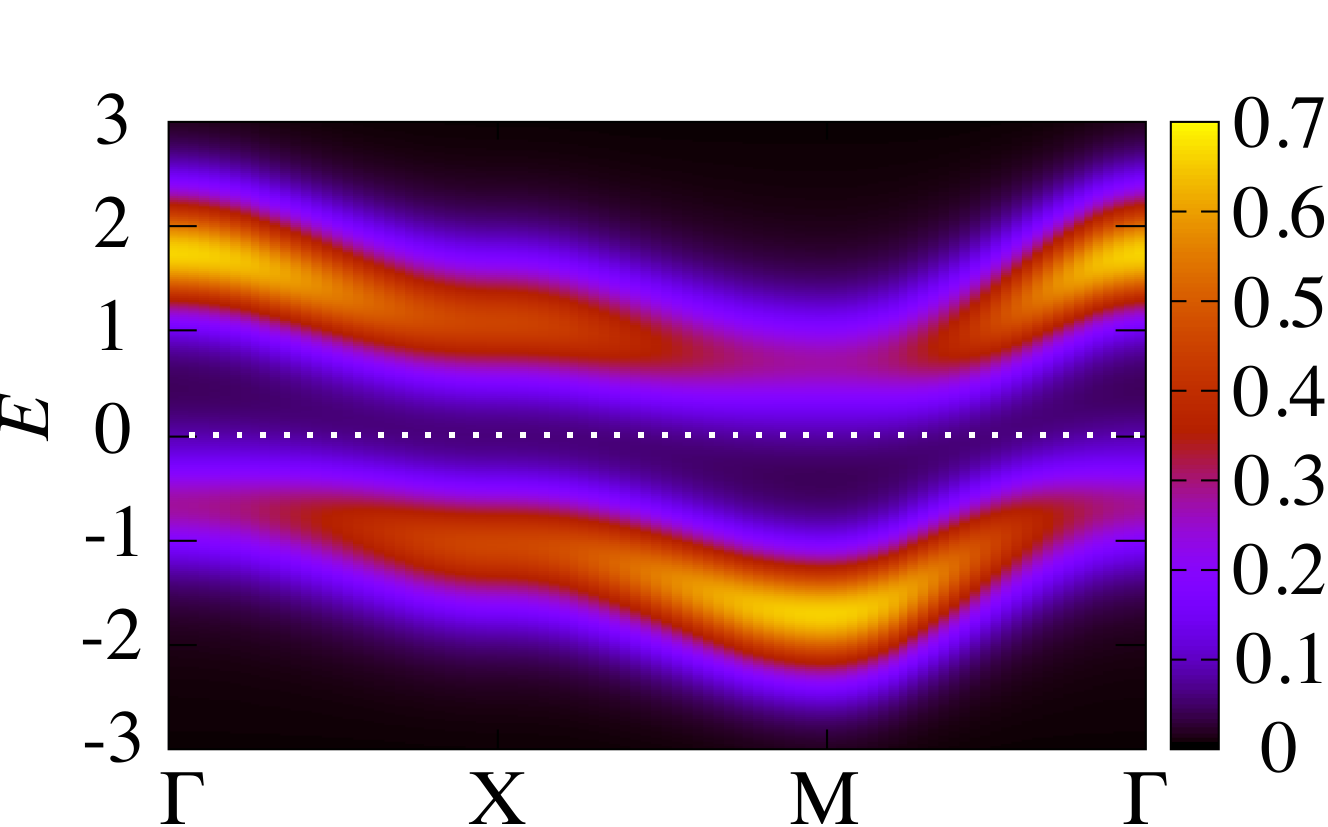}
~~
\includegraphics[width=0.19\linewidth]{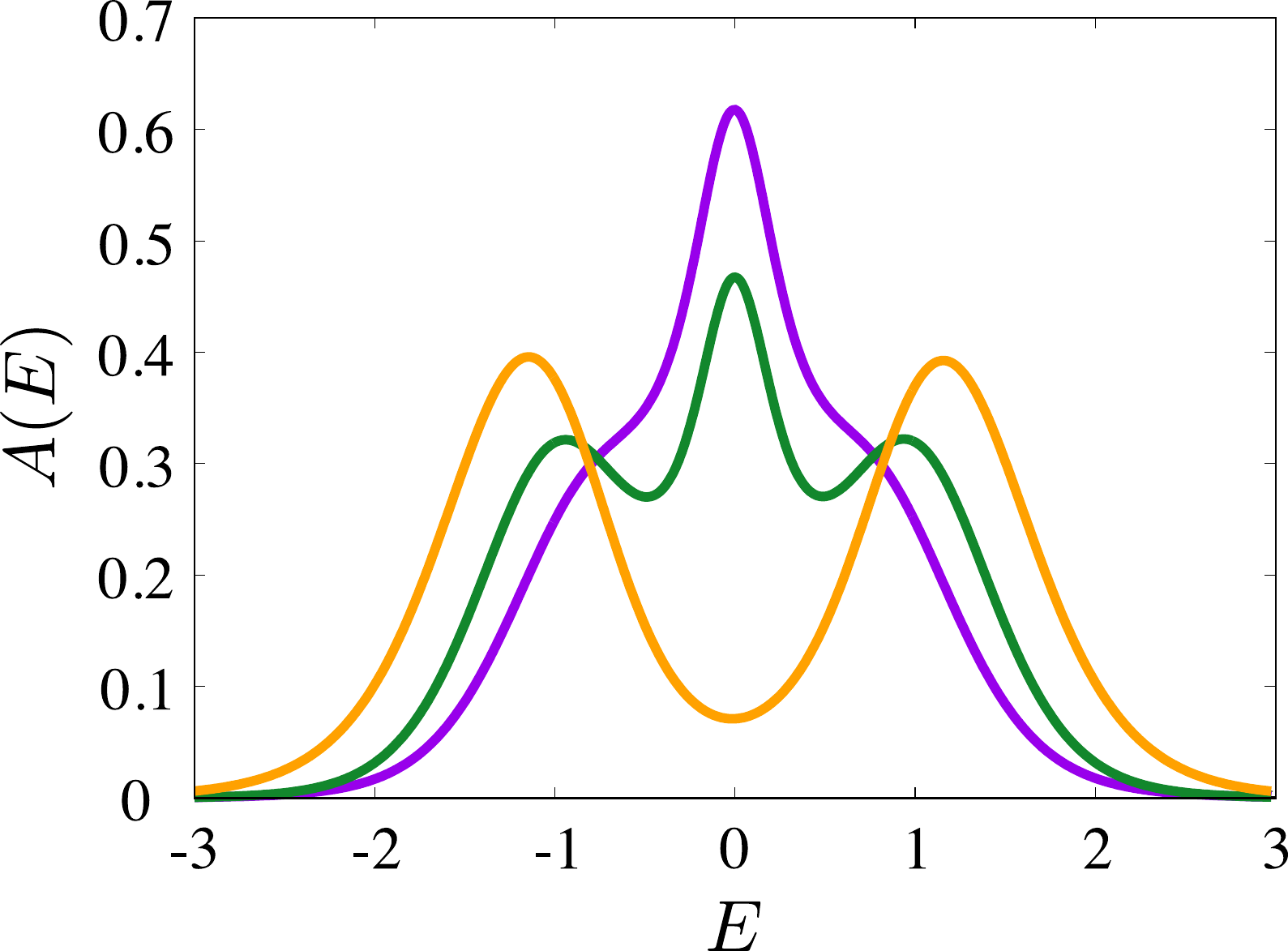}
\\
\includegraphics[width=0.25\linewidth]{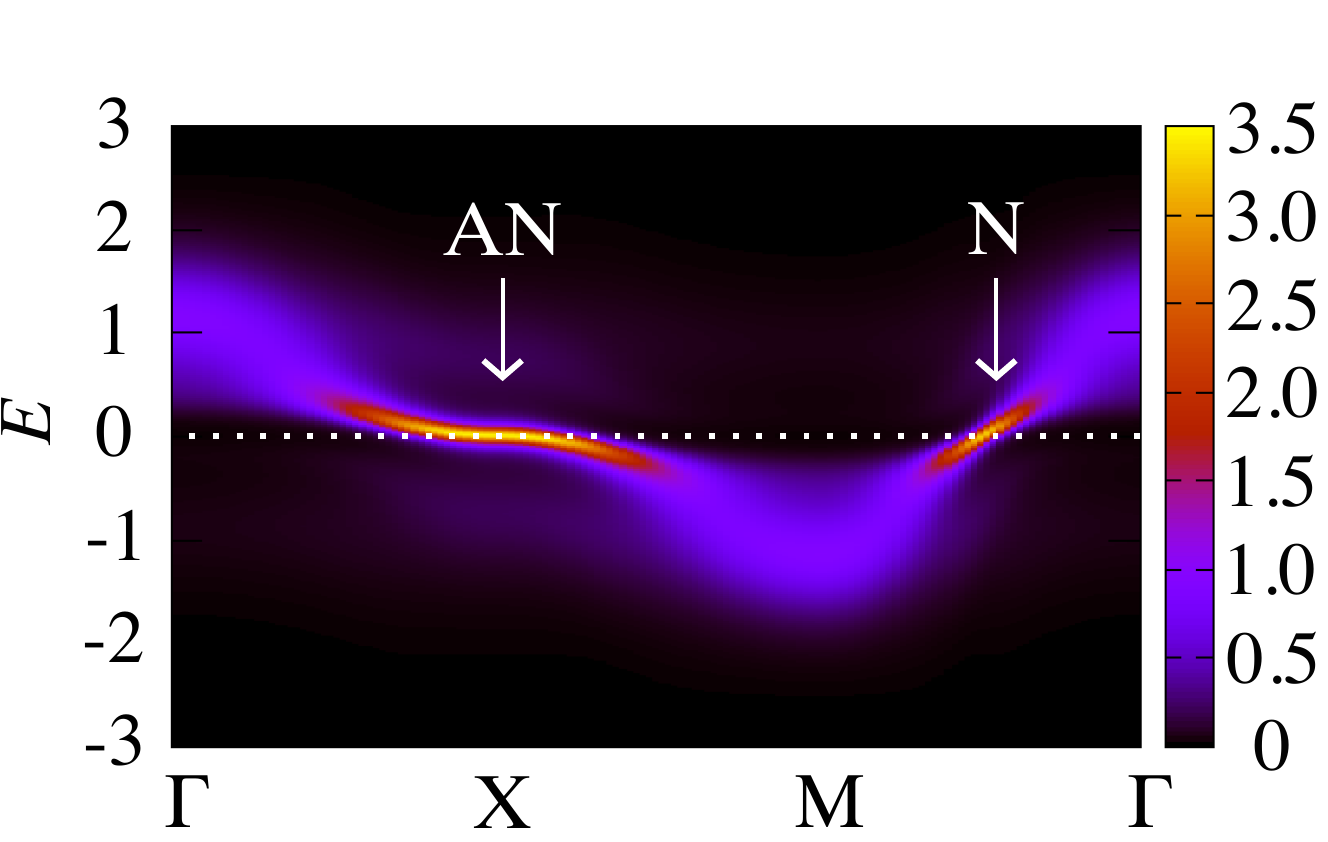}
~~
\includegraphics[width=0.25\linewidth]{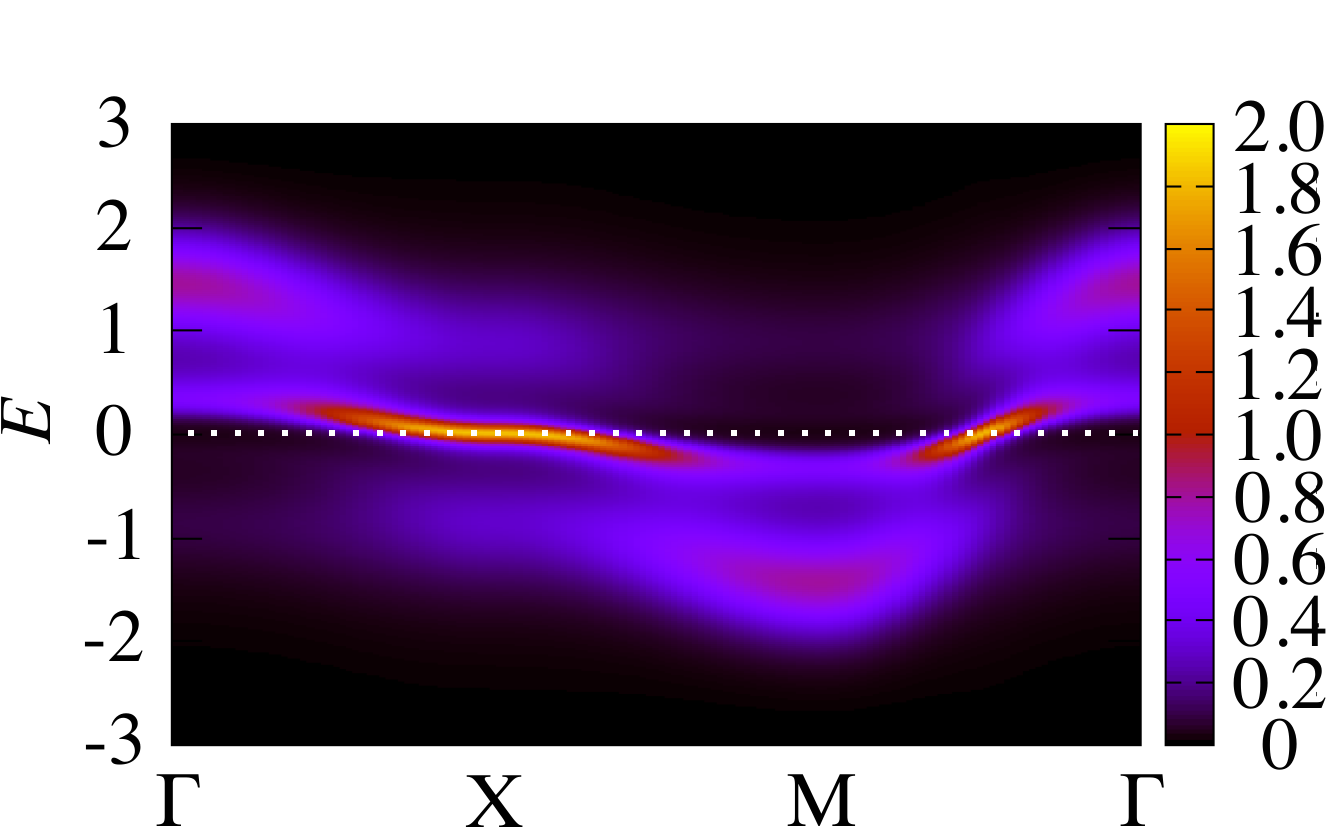}
~~
\includegraphics[width=0.25\linewidth]
{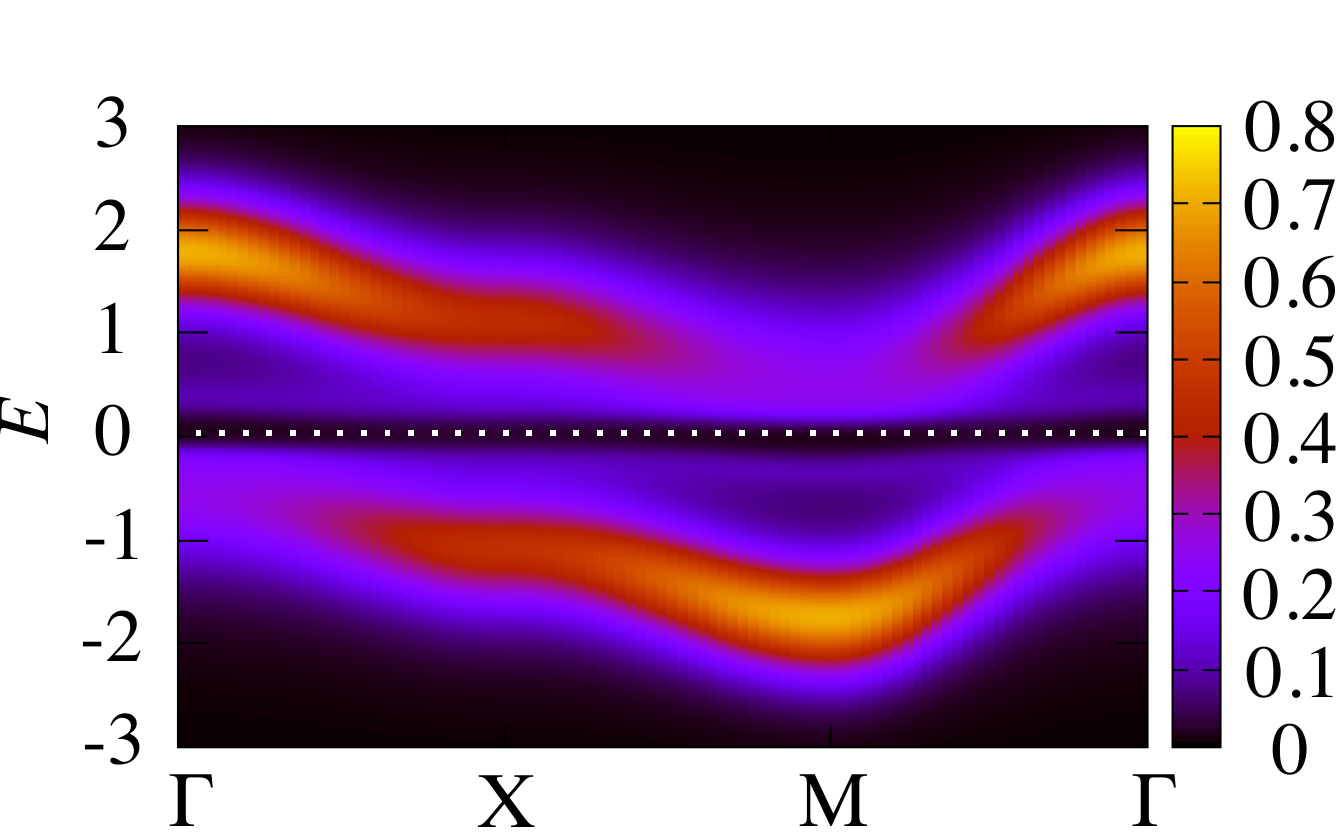}
~~
\includegraphics[width=0.19\linewidth]{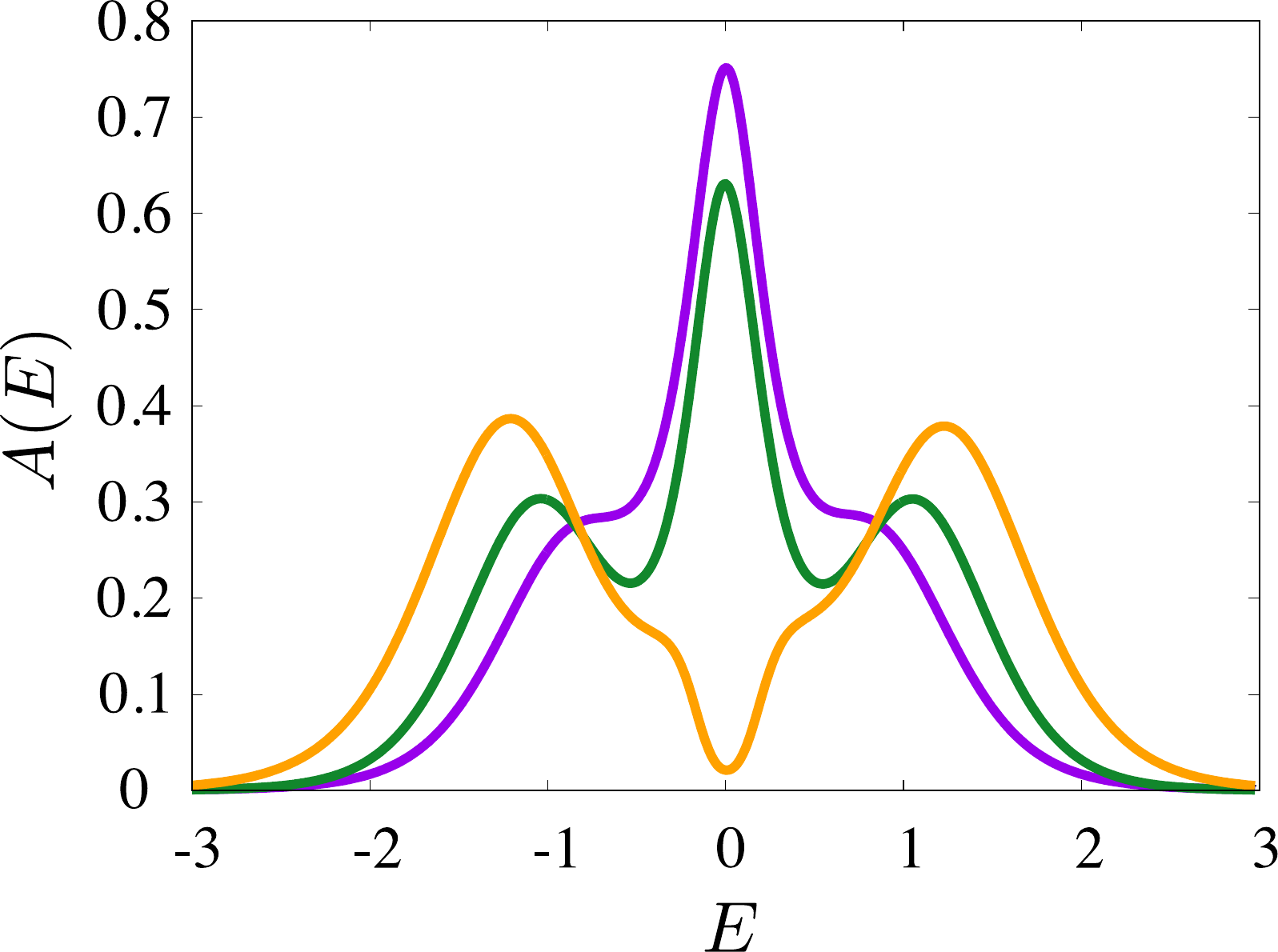}
\caption{Momentum-resolved spectral function of the single-band Hubbard model at half filling calculated with single-site DMFT, for two temperature regimes ${T=0.14}$ (upper row) and ${T=0.06}$ (lower row). 
From left to right the spectral function is plotted for interactions $U=1.2$, $U=1.6$, and $U=2.4$. The far right column displays the corresponding local spectral functions obtained for ${U=1.2}$ (purple), ${U=1.6}$ (green) and ${U=2.4}$ (orange).} 
\label{fig:spectr_highT_dmft_dtrilex}
\end{figure}

\twocolumngrid

\section{I. Momentum-resolved spectral function in DMFT}

In Fig.~\ref{fig:spectr_highT_dmft_dtrilex} we present results of the momentum-resolved spectral function calculated within DMFT, allowing for a comparison with the D-TRILEX results shown in Fig.~1 of the main text. We observe that at high ${T=0.14}$ (upper row of Fig.~\ref{fig:spectr_highT_dmft_dtrilex}) the results are in very good agreement with \mbox{D-TRILEX}, since here the magnetic fluctuations are weak and a local picture describes accurately the system. However, at low ${T=0.06}$ (lower row of Fig.~\ref{fig:spectr_highT_dmft_dtrilex}), the DMFT results are qualitatively similar to the high $T$ ones, in contrast to the findings of D-TRILEX, where strong spatial fluctuations completely alter the systems behavior, as discussed in detail in the main text.

\section{II. Energy spectrum near a magnetic instability}
\label{app:Mirroring}

\begin{figure*}
\includegraphics[width=1\linewidth]{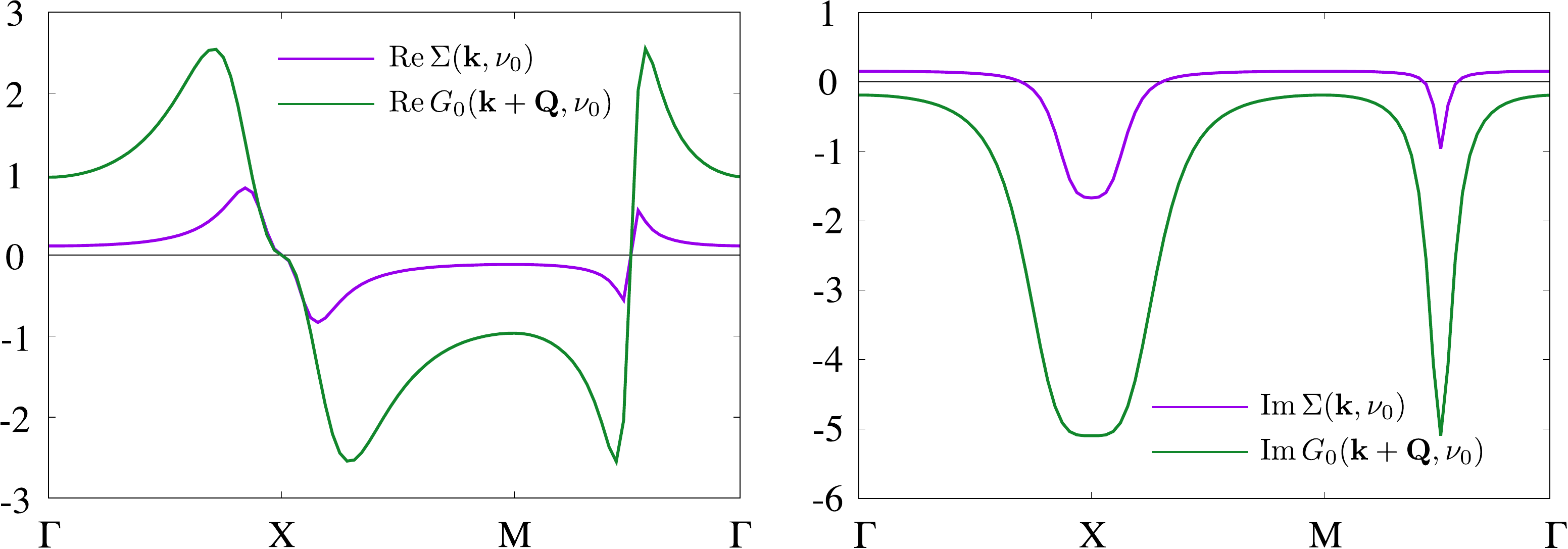}
\caption{Real (left panel) and imaginary (right panel) parts of the lattice self-energy $\Sigma({\bf k},\nu)$ and of the of the bare lattice Green's function $G_0({\bf k+Q},\nu)$. Calculations for the self-energy are performed using \mbox{D-TRILEX} for ${U=1.1}$, ${T=0.06}$ and for the zeroth Matsubara frequency $\nu_0$ along the high-symmetry path in the BZ. The two function are indeed proportional, following the expression~\eqref{eq:dualself}.
\label{fig:appendix}}
\end{figure*}

As discussed in the main text, at low temperatures we observe a mirroring effect of the electronic dispersion with respect to zero energy. In the weak coupling regime, we can understand the emergence of this effect in the following way. Indeed, in that regime one can approximate the lattice self-energy by a $GW$-like form in the spirit of the FLEX~\cite{BICKERS1989206, PhysRevB.57.6884} or \mbox{D-TRILEX}~\cite{PhysRevB.100.205115, PhysRevB.103.245123, 10.21468/SciPostPhys.13.2.036} approaches:
\begin{align}
\Sigma({\bf k},\nu) = \frac{1}{N\beta}\sum_{{\bf q},\omega,\varsigma} G_{0}({\bf k+q},\nu+\omega) W^{\varsigma}({\bf q},\omega),
\end{align}
where ${G_{0}({\bf k},\nu)=\left[i\nu-\varepsilon_{\bf k}\right]^{-1}}$ is the bare lattice Green's function and ${W^{\varsigma}({\bf q},\omega)}$ is the renormalized interaction in the charge ${\varsigma=c}$ and spin ${\varsigma=s\in\{x,y,z\}}$ channels.
In our notations, $W^{\varsigma}$ is positive: ${W^{\varsigma}\geq0}$.
$N$ is the number of ${\bf k}$-points in the discretized BZ, $\varepsilon_{\bf k}$ is the electronic dispersion, $\beta$ is the inverse temperature, and $\nu$ ($\omega$) is a fermionic (bosonic) Matsubara frequency. 
In the vicinity of the magnetic instability the main contribution to the self energy is given by the renormalized interaction in the spin channel.
In this regime the latter has a delta-function-like form ${W^{s}({\bf q},\omega)\simeq{}W^{s}\delta_{{\bf q,Q}}\delta_{\omega,0}}$, where the momentum ${\bf Q}$ corresponds to the ordering vector.
Under this assumption the self-energy simplifies to:
\begin{align}\label{eq:dualself}
\Sigma({\bf k},\nu) \simeq \frac{3W^{s}}{N\beta} G_{0}({\bf k+Q},\nu).
\end{align}
The energy spectrum is defined by the poles of the dressed Green's function $G({\bf k},\nu)$ that can be found through the Dyson equation:
\begin{align}
G^{-1}_{0}({\bf k},\nu) = G^{-1}_{0}({\bf k},\nu) - \Sigma({\bf k},\nu),
\end{align}
which gives the following expression for the energy ${E=i\nu}$:
\begin{align}
E^{2} - E \left(\varepsilon_{\bf k} + \varepsilon_{\bf k+Q} \right) + \varepsilon_{\bf k}\varepsilon_{\bf k+Q} - \frac{3W^{s}}{N\beta}=0.
\end{align}
If the last term in this expression is small, the energy spectrum is given by the two quasiparticle bands:
\begin{align}
E_{1} = \varepsilon_{\bf k} ~~~ \text{and} ~~~ E_{2} = \varepsilon_{\bf k+Q}.
\label{eq:en_sp}
\end{align}
In the case of a square lattice with nearest-neighbor hopping amplitudes the ordering vector corresponds to AFM fluctuations ${{\bf Q}=(\pi,\pi)}$, and the dispersion relation satisfies ${\varepsilon_{\bf k+Q} = - \varepsilon_{\bf k}}$.
Consequently, the energy spectrum reveals two mirrored quasiparticle bands that are visible in the bottom left and middle panels of Fig.1 in the main text.
In fact, Fig.~\ref{fig:appendix} shows that the lattice self-energy $\Sigma({\bf k},\nu)$ obtained using \mbox{D-TRILEX} in the vicinity of the AFM instability is proportional to the bare lattice Green's function $G_0({\bf k+Q},\nu)$ in agreement with Eq.~\eqref{eq:dualself}, which eventually leads to the energy spectrum~\eqref{eq:en_sp}.

\section{III. Formation of the local magnetic moment}

\begin{figure}[b!]
\includegraphics[width=\linewidth]{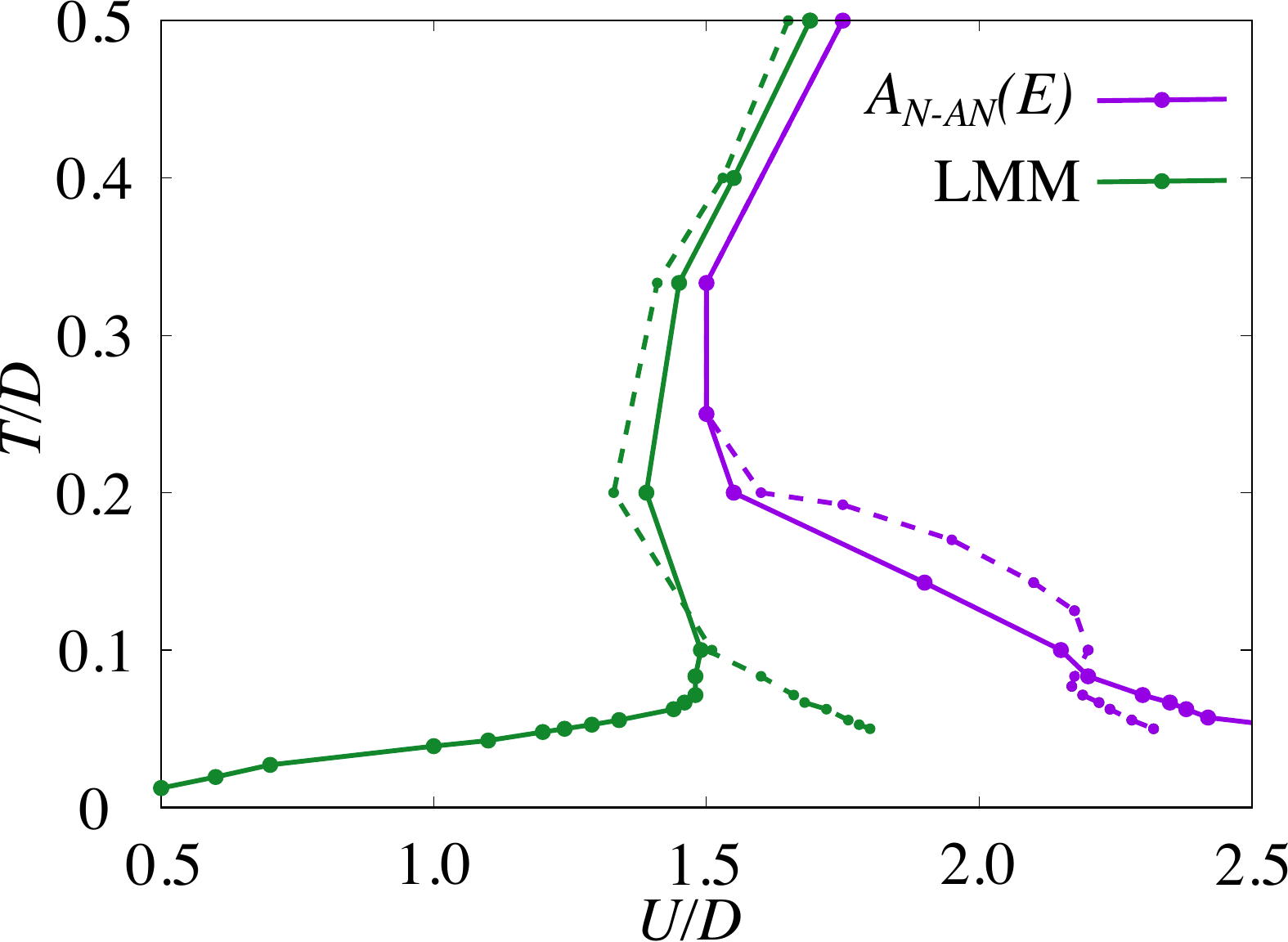}
\caption{Formation of a minimum at the $E_{F}$ at the AN and N points (purple) calculated with \mbox{D-TRILEX}. The formation of the LMM is shown in green. The DMFT results are added in thin lines for comparison. At large $T$ DMFT and \mbox{D-TRILEX} results are nearly identical. Decreasing $T$ and consequently increasing the strength of magnetic fluctuations leads to a back-turn in the LMM in \mbox{D-TRILEX}. The Mott transition line, identified from the simultaneous formation of the minimum at $E_{F}$ at N and AN points (purple), is found at slightly larger values of $U$ compared to DMFT.}
    \label{fig:phasediag_largeT_DTRILEX}
\end{figure}

\begin{figure*}
\includegraphics[width=1\linewidth]{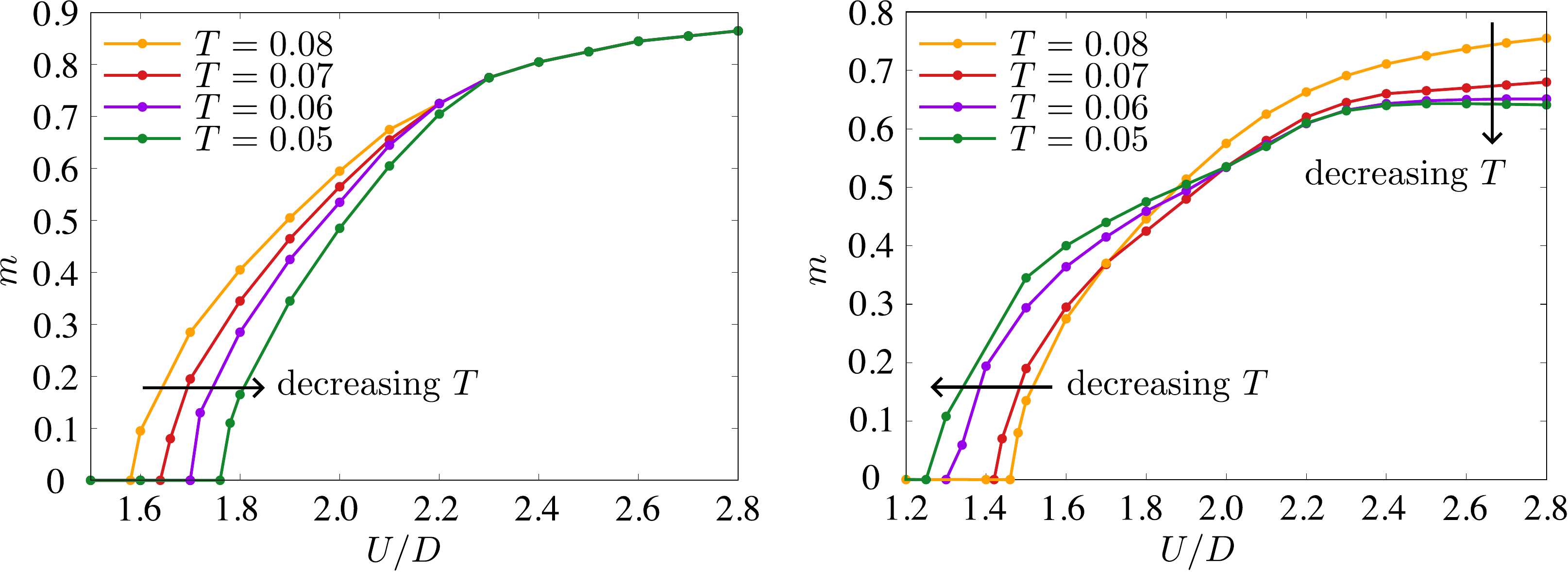}
\caption{The value of the local magnetic moment $m$ as a function of the interaction strength for different values of temperature calculated using DMFT (left panel) and \mbox{D-TRILEX} (right panel). Within DMFT the formation of the LMM appears at larger $U$ for a decreasing $T$ and all the curves saturate to a value close to $1$ once the Mott transition takes place. 
On the contrary, in D-TRILEX the LMM is formed at larger $U$ for an \textit{increasing} $T$ and the order of the different curves is reversed at intermediate interactions with the lines related to the higher $T$ saturating at larger values compared to the low-$T$ ones. The LMM is now screened due to the magnetic fluctuations which are stronger at smaller $T$ leading to the relevant saturation to a smaller value.
\label{fig:locmom_dmft_dtrilex}}
\end{figure*}

In the main text we discuss the formation of the local magnetic moment in the system. In this section we analyze this effect in detail by performing calculations with both DMFT and \mbox{D-TRILEX} at a large temperature range. 
In Fig.~\ref{fig:phasediag_largeT_DTRILEX} we plot the green curve corresponding to the formation of the local magnetic moment (LMM) in the system, with solid line for \mbox{D-TRILEX} and with dashed line for DMFT. 
As has been shown in Ref.~\onlinecite{PhysRevB.105.155151}, the formation of the LMM occurs via a spontaneous symmetry breaking in the local free energy of the system (with excluded contribution of itinerant electrons) as a function of the magnetic moment $m$. 
The symmetry breaking can be seen in the change of the form of the free energy from a parabola-like with a minimum at ${m = 0}$ to a double-well potential with two minima at ${m \neq0}$.
The green curves in Fig.~\ref{fig:phasediag_largeT_DTRILEX} have been extracted from calculations like those shown in Fig.~\ref{fig:locmom_dmft_dtrilex} (left panel for DMFT and right panel for D-TRILEX), by identifying for each $T$ the critical $U$ at which the LMM is formed. 
This curve, thus, separates the Slater (weak coupling) regime of electronic correlations with no LMM from the Heisenberg (strong coupling) regime with the formed LMM~\cite{PhysRevB.105.155151}. 
The purple lines depict the critical interactions and temperatures at which the quasiparticle peak at $E_{F}$ disappears {\it simultaneously} at the AN and N points of the FS.
The purple lines are thus associated with the Mott transition. 
As could be expected, at large $T$ the two methods (DMFT and D-TRILEX) give nearly identical results, since in this regime the nonlocal fluctuations are very weak and DMFT is already enough for an accurate description of the system. At this regime the green curves lie to the left of the purple lines and have a nearly identical to them shape thus acting as a forerunner of the MIT to a Mott state, with a turn above ${T\simeq0.2}$. 
At ${T<0.1}$, when the spatial magnetic fluctuations start to play a significant role (${\lambda>0.65}$), a qualitative difference between the two methods is observed. 
We find, that for DMFT at low temperatures the LMM curve lies approximately parallel to the metal-to-Mott-insulator phase boundary (purple line) but is shifted to smaller interaction strengths,  acting, as for large $T$, as a precursor of the formation of the Mott state.
The bold purple line, that corresponds to a \mbox{D-TRILEX} prediction for a simultaneous disappearance of the quasiparticle peak along the FS, follows the DMFT trend (thin purple line) toward stronger interactions, and is even pushed to slightly larger values of $U$. 
The bold green curve depicting the LMM formation in \mbox{D-TRILEX}, however, turns to smaller values of $U$, departing from its DMFT counterpart (thin green line) at ${T<0.1}$.

\begin{figure*}[t!]
\includegraphics[width=0.48\linewidth]{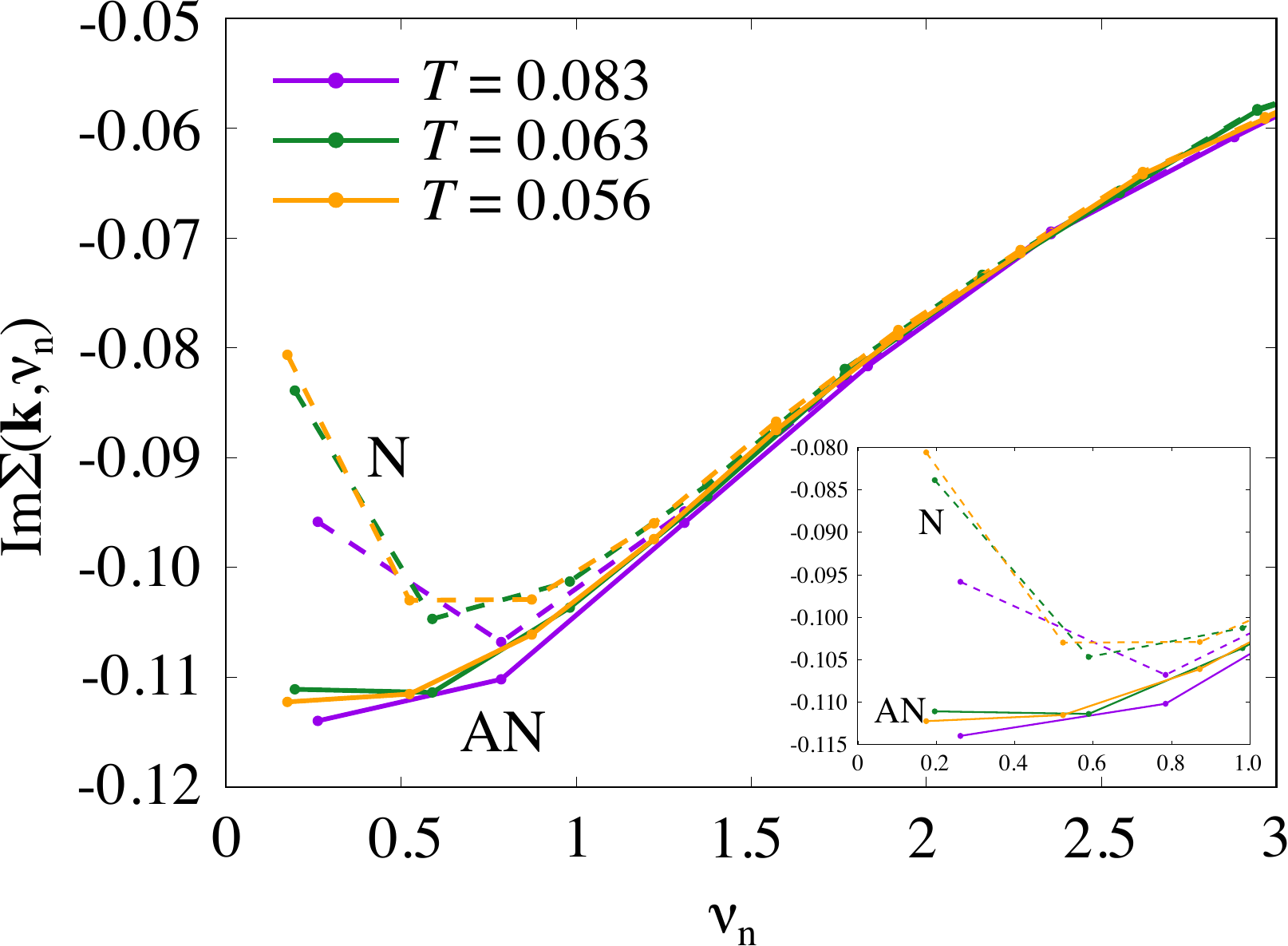}  ~~~~
\includegraphics[width=0.48\linewidth]{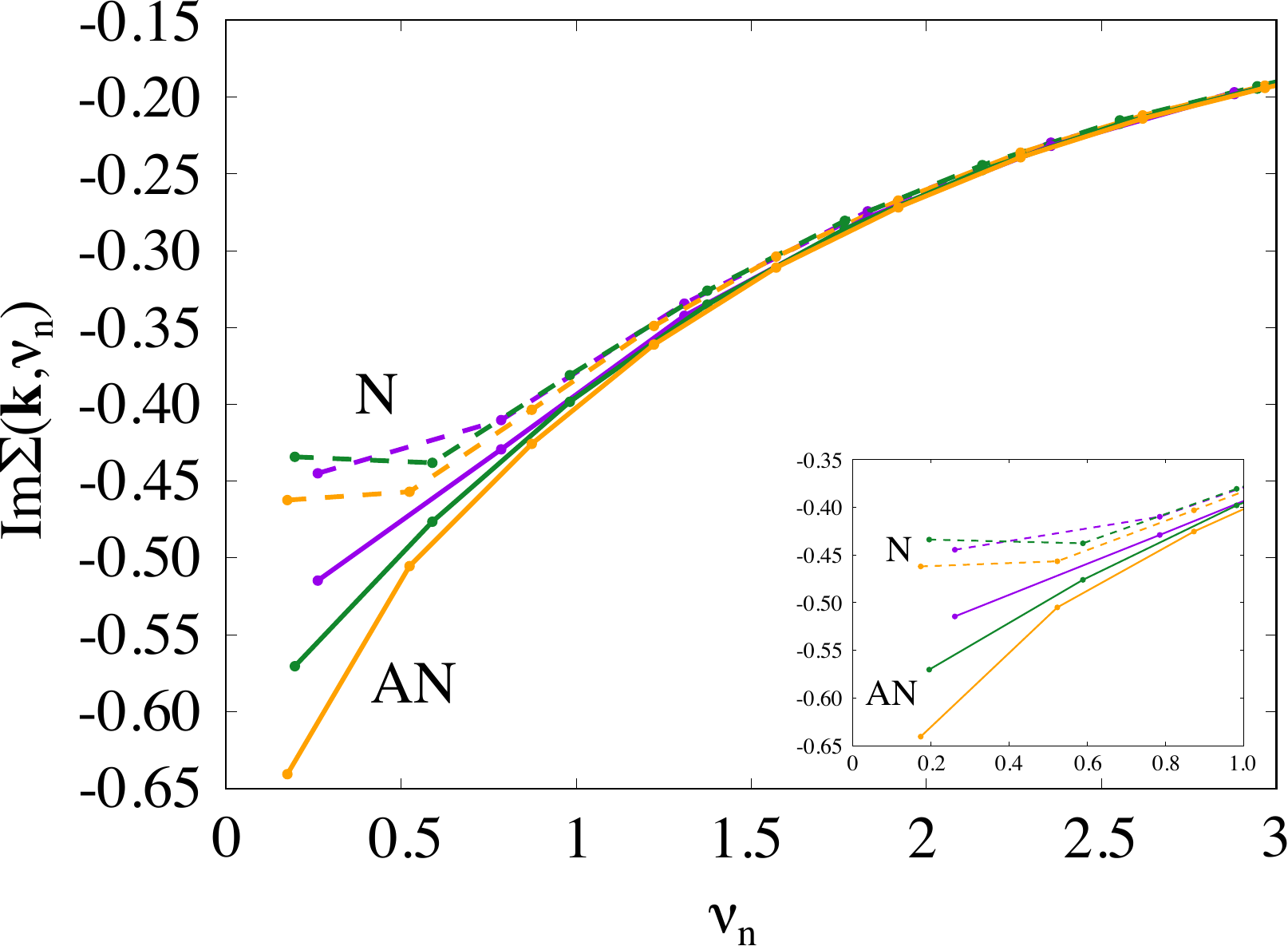}
\caption{Imaginary part of the self-energy obtained for different temperatures ${T=0.083}$, ${T=0.063}$, and ${T=0.056}$. Results are shown for the AN (solid lines) and N (dashed lines) points for two different values of the interaction ${U=0.9}$ (left panel) and ${U=1.6}$ (right panel). 
In the two insets the results are plotted within the range of the two first Matsubara self-energies, the difference of which is used in determining the nature of the system as a Fermi liquid (FL) one ${\left({\rm Im}\,\Sigma(i\omega_0)-{\rm Im}\,\Sigma(i\omega_1)>0\right)}$ or a non-Fermi liquid (NFL) one (eventually becoming an insulator) ${\left({\rm Im}\,\Sigma(i\omega_0)-{\rm Im}\,\Sigma(i\omega_1)<0\right)}$. 
For the AN point at ${U=0.9}$ and for the N point at ${U=1.6}$ the self-energy shows a FL behavior at intermediate $T$ and a NFL behavior at lower and higher $T$. Such results give rise to the behavior of the self-energy (light red and light blue) curves shown in Fig.~2 of the main text.
\label{fig:selfen_N_AN}}
\end{figure*}

We can now look back at Fig.~\ref{fig:locmom_dmft_dtrilex} and use the behavior of the value of the LMM in order to explain how the crossover regime at intermediate interactions transforms to a strong-coupling Mott regime governed by the local electronic correlations. 
As already introduced, in the left panel of Fig.~\ref{fig:locmom_dmft_dtrilex} we plot the value of the LMM $m$ predicted by paramagnetic DMFT calculations as a function of interaction strength $U$ for different temperatures. 
We find, that decreasing the temperature decreases accordingly the critical interaction for the formation of the LMM, as seen in the low-$T$ part of Fig.~\ref{fig:phasediag_largeT_DTRILEX}, with the thin green line. 
By further increasing $U$ the different curves approach each other until they eventually merge successively into one line. 
It is interesting to point out, that if we trace the merging of each curve with the main one we obtain the Mott transition line plotted with the thin purple curve in Fig.~\ref{fig:phasediag_largeT_DTRILEX}. 
Therefore, the saturation of the value of the LMM (which tends toward the value $1$ at large $U$) can be used as a second indication of the DMFT Mott transition.

Similarly, the right panel of Fig.~\ref{fig:locmom_dmft_dtrilex} illustrates the same calculations done now within \mbox{D-TRILEX}. 
We here observe a different behavior compared to DMFT. 
In particular, we find that instead of merging into one line, the LMM curves saturate to different values of $m$ upon increasing the interaction. 
This can be understood considering that nonlocal magnetic fluctuations prohibit the LMM from being completely formed, thus the proximity to the magnetic instability (tuned by the temperature) results into a screening of the value of the LMM.
However, surprisingly, we observe that the value of the LMM ${m=0.65}$, which corresponds to the saturation of the lowest $T$ above $T_N$, crosses each temperature curve in the right panel of Fig.~\ref{fig:locmom_dmft_dtrilex} at the $U_c$ that is in a very good agreement with the critical interaction for the Mott transition predicted by \mbox{D-TRILEX}.
This result suggests that the end of the crossover regime occurs upon reaching the critical value of the LMM effectively when the system undergoes the Mott transition.

\section{IV. Behavior of the self-energy}

As discussed in the main text, a signature of the suppression of the spatial electronic fluctuations can be found by investigating the behavior of the electronic self-energy.
There, the signatures of the MIT can be obtained by looking at the imaginary part of the self-energy (${{\rm Im}\,\Sigma({\bf k},\nu_n)}$) at AN and N points as a function of fermionic Matsubara frequencies $\nu_n$, such as those shown in Fig.~\ref{fig:selfen_N_AN} for two values of interaction and for a temperature scan. 
In order to determine the nature of the system through the self-energy we compare its value at the two first frequencies. 
When ${{\rm Im}\,\Sigma(\nu_0)-{\rm Im}\,\Sigma(\nu_1)>0}$ we denote the system as a Fermi liquid (FL) 
\footnote{This is a somewhat loose definition. For true quasi-particle excitations in the sense of
Landau to exist, we would need to also check the value of the zero-frequency intercept indicating the
inverse lifetime of the quasi-particle.}
and when ${{\rm Im}\,\Sigma(\nu_0)-{\rm Im}\,\Sigma(\nu_1)<0}$ it has lost the FL coherence (which for simplicity, we denote as NFL), eventually evolving to an insulator. 
For both the N and AN points we find a non-monotonous behavior as a function of $T$, however taking place at smaller values of the interaction for the AN point (e.g. ${U=0.9}$, left panel) compared to the N one (e.g. ${U=1.6}$, right panel). 
At ${U=0.9}$, while the self-energy at the N point exhibits a metallic FL behavior for all the temperatures shown, the self-energy at the AN is NFL for the lowest and highest $T$ and is FL at the intermediate $T$. Similarly, for ${U=1.6}$ the self-energy at the AN point now displays a NFL behavior for all temperatures, while the self-energy at the N point is NFL at low and high $T$, and is FL at the intermediate $T$. Such detailed calculations for different values of $U$ and $T$ give rise to the self-energy curves plotted in the phase diagram, Fig.2 of the main text.

\bibliography{supp}